\documentclass{aa}  
\usepackage{graphicx} 
\usepackage{booktabs}
\usepackage[table,xcdraw]{xcolor}
\usepackage[version=3]{mhchem}
\usepackage{colortbl}

\usepackage{txfonts}
\usepackage[table]{xcolor}
\usepackage{subcaption}
\usepackage{textcomp}
\usepackage[export]{adjustbox}
\usepackage{booktabs}
\usepackage{natbib}
\bibpunct{(}{)}{;}{a}{}{,}
\usepackage{txfonts}
\usepackage[colorlinks,citecolor=blue,linkcolor=red,urlcolor=cyan]{hyperref}

\newcommand{\teff}  {T$_\mathrm{eff}$}

\newcommand{\kms}{\hbox{km/s}}
\newcommand{\ha}  {H$\alpha$}
\newcommand{\has}  {H$\alpha$06}
\newcommand{\hal}  {H$\alpha$16}

\begin{document}

   \title{The enigma of Li-rich giants and its relation with temporal variations observed in radial velocity and stellar activity signals}

   \author{Inês Rolo
          \inst{1,2},
          Elisa Delgado Mena\inst{1},
          Maria Tsantaki \inst{3},
          João Gomes da Silva\inst{1}
          }

   \institute{Instituto de Astrofísica e Ciências do Espaço, Universidade do Porto, CAUP, Rua das Estrelas, 4150-762 Porto, Portugal (\email{ines.rolo@astro.up.pt})
         \and
             Departamento de Física e Astronomia, Faculdade de Ciências, Universidade do Porto, Rua do Campo Alegre 687, 4169-007 Porto, Portugal
         \and
            INAF – Osservatorio Astrofisico di Arcetri, Largo E. Fermi 5, 50125 Firenze, Italy
             }

   \date{Received Date / Accepted Date}

 
  \abstract
   {{Despite the large number of studies focused on the characterisation of Li-rich stars and understanding the mechanisms leading to such enrichment, their origin remains a mystery.}}
   {Magnetic activity, particularly the phenomena usually associated with it (e.g. spots and plages), and the Li abundance (A(Li)) of stars, are in general thought to be connected. As of today, however, just how they are connected is unclear. In this work, we study a sample of young but evolved intermediate-mass red giants that are inhabitants of open clusters where planets have been searched for. Our aim is to use radial velocity (RV) and stellar activity indicator signals to look for relations between Li abundances and stellar activity or variability.}
   {We explored how the standard deviation (STD), peak-to-peak amplitude (PTP), mean, and median of typical stellar activity indicators (BIS, FWHM, $T_\textrm{eff}$, and H$\alpha$ index) change as a function of the Li content of 82 red giants. Furthermore, we computed weighted Pearson correlation coefficients ($\rho_w$) between time series of RV measurements and the stellar activity indicators for the stars in our sample. To aid our results, we also studied generalized Lomb-Scargle periodograms (GLSP) to capture possible significant periodic temporal variations in our data.}
   {Our analysis indicates that the STD and PTP of BIS and FWHM, the mean and median of the \ha\ index, and $v\sin(i)$ increase exponentially with A(Li) in our sample of red giants. Significant temporal variations and correlations between RVs and activity indicators also tend to be found preferentially for stars where high A(Li) is observed. Most of the Li-rich stars in our sample either show strong correlations {of RV} with at least one of the stellar activity indicators or reveal significant periodic temporal variations in their GLSPs of stellar activity indicators that are consistent with those found for RV.}
   {}

   \keywords{red-giants -- open clusters -- lithium rich -- stellar activity
               }

\titlerunning{The enigma of Li-rich giants and its relation with stellar activity}
\authorrunning{Rolo et al.}
\maketitle

%

\section{Introduction}

Before entering the red giant branch (RGB) of their evolution, low- ($\lesssim2M_\odot$) to intermediate-mass stars ($\sim 2-8M_\odot$) {\citep{1967bIben,1967aIben}} go through a mixing episode called the first dredge up (FDU), which destroys most, if not all, of their surface lithium (Li). Stars as massive as the Sun, according to standard stellar evolution models, should not exhibit a surface lithium abundance, A(Li)\footnote{${\rm A(Li)}=\log[N({\rm Li})/N({\rm H})]+12$}, exceeding 1.5 dex after the FDU. In general, observations concur with this, but $\sim$1\% of observed giants {(e.g. \citealt{Brown1989,2011Kumar,2014Liu,2016Casey,Kirby_2016,2019Deepak,Gao2019,2020Pabst,2021Martell})} present abnormally high A(Li), with the highest abundance observed to date in a red giant star being 5.62 dex \citep{Jeremy2022}. These stars are known as Li-rich giants. 

Possible explanations for the Li enrichment observed in these stars fall under two main categories. First, it has been proposed that a star may show high levels of surface Li if it has been contaminated by an external source. For example, if planet engulfment occurs, the Li content of the planet can be laid out in the photosphere of the star and enrich it (e.g. {\citealt{Adamow2012,Aguilera2016,mena2016searching}}). On the other hand, under certain conditions, Li can be synthesised inside a star and rise to the surface through extra mixing processes. For stars with mass $>4 M_\odot$ in the asymptotic giant branch (AGB) of their evolution, the Cameron-Fowler mechanism (CFM; \citep{CameronFowler1971}) is the preferred scenario to explain Li enrichment. In these stars, the temperature at the bottom of the convective envelope is favourable to the production of beryllium, which when transported fast enough to the outer layers of the star can produce Li without it being destroyed. In lower-mass stars, high enough temperatures for beryllium production can only be found inwards of the convective envelope. For these stars, the existence of extra mixing mechanisms such as thermohaline mixing (e.g. \citealt{Egg2008,Cantiello2010}) and rotation-induced mixing {\citep[e.g.][]{2004Denissenkov,Palacios2004,Adamow2014}} are more plausible, as they are needed to bring the beryllium from the deeper layers to the convective envelope so that CFM can ensue. One way to probe extra mixing scenarios is by analysing the surface carbon isotopic ratio \ce{^{12}C/^{13}C}, which after the FDU occurs is lowered to values of about 20-30 {\citep{1994Charbonnel,2004Denissenkov}}. Further lowering of \ce{^{12}C/^{13}C} is a good and widely used indication that extra mixing occurred in a star {\citep{Charbonnel1995,2023Aguilera,2024A&Lagarde}}.  

It is also generally accepted that higher than normal A(Li) is common among magnetically active stars, even if the reason for this is still poorly understood. This has been shown in different types of stars, including dwarfs \citep{favata1996lithium,Zboril1997dwarfs}; magnetically peculiar stars (see \citealt{lyubimkov2016lithium} for a review); active binaries and pre-main-sequence stars \citep{pallavicini1992lithium}; active main-sequence and zero-age main-sequence stars \citep{Xing2021}; and {active giant stars} \citep{fekel1994lithium,sneden22}. 

In this work, we are interested in studying the possible connection between Li enhancement in red giant stars and their stellar activity. {As such, we started by looking for relations between several descriptive statistics of stellar activity indicators and A(Li) in a sample of red giant stars, some of which are lithium rich, where planets have been searched. These stars are inhabitants of open clusters, which is particularly favourable, as we are able to better constrain their evolutionary status. To strengthen our analysis, we also looked for correlations as well as periodic temporal variations in stellar activity signals that may connect the variability observed in radial velocity (RV) to stellar activity.} The present article is organised as follows: In Sect. \ref{obs}, we provide an overview of our data. In Sect. \ref{methods}, we explain the methods incurred in the data analysis. In Sect. \ref{discres}, we discuss our results, and the main conclusions are laid out in Sect. \ref{conc}.

\section{Observations and sample}
\label{obs}

Our sample is composed of young but evolved intermediate-mass red giants residing in ten different open clusters (IC4651, IC4756, NGC2345, NGC2423, NGC3114, NGC3532, NGC3680, NGC4349, NGC5822, and NGC6705). These stars have been observed over the years with the HARPS spectrograph \citep{Mayor2003} at the ESO 3.6-metre telescope (R=115\,000) and the CORALIE spectrograph \citep{Queloz2000,Udry2000} at the Swiss 1.2-metre Leonhard Euler Telescope at ESO's La Silla Observatory (R=60\,000) as part of a programme to search for exoplanets. (For a complete description of the RV surveys see \citet{LovisMayor2007,ElisaPlanets,Elisa2023}.) The stellar parameters and Li abundances for this sample are presented in \cite{MariaLi}, who used the same spectra employed here.

In this work, the data available from CORALIE are not treated. CORALIE data carry a much higher associated uncertainty in the measurements than HARPS data due to a significantly lower resolution (R). Furthermore, CORALIE data were only collected for a subset of stars in our sample, and the
measurements for each star are few in number. For these reasons
we preferred to use only HARPS data.

Only stars with five or more observational data points were considered for analysis. This limitation increases the credibility of our results while including the most observed stars as possible. In total, 82 stars were analysed out of which 15 are canonically Li-rich, that is they show A(Li)$>$1.5 dex. If a Population I star forming with a metallicity similar to that of the Sun has in the beginning of its life A(Li)$\sim$3.3 dex, that is, the approximate A(Li) of the interstellar medium, accounting for the dilution occurring during the FDU, it should be left with a maximum of A(Li)$\sim$1.5 dex in the red giant phase of its evolution (e.g. \citealp{charbonnel2000nature}). Studying stars in open clusters, however, allows for a more precise estimate of this limit. Open clusters are ideal laboratories for studying Li enhancement given that they provide natural constraints on the mass, composition, age, and evolutionary state of their members. {With such information, it is possible to estimate the amount of Li that all stars in a given cluster should have at their current age using stellar evolution models. This allows us to better constrain what makes {a star in a given cluster `Li-rich'} (see \citealt{Sun2022}).} \cite{MariaLi} conducted such a study, providing estimations for the maximum A(Li) that stars in 32 different open clusters should have in their current evolutionary stage. {Stars are considered Li-rich according to the limits defined in Table \ref{table:li_limits} and the canonical definition.} Special attention is given to stars that show high Li content in comparison to inhabitants of the same cluster in the same evolutionary stage. We deem such stars `Li-Rich-in-Context', in accordance with \cite{MariaLi} and \cite{Sun2022}. All relevant stars are presented in Table \ref{table:lirich}. Some of these stars have been studied previously in the literature. A thorough discussion concerning the possible nature of the Li enrichment observed in these stars has been provided in \citet{MariaLi}, and it is summarised in Table \ref{table:lirich}.

\begin{table}[h!]
\centering
\caption{Limits over which a star can be considered lithium rich for every cluster in our sample (A(Li$)_{\text{max}}$[dex]). Also included is the metallicity of each cluster ([Fe/H]).}
\begin{tabular}{@{}ccc@{}}
\toprule
Cluster & A(Li$)_{\text{max}}$[dex] & [Fe/H] [dex] \\ \midrule
IC4651  & 1.46                       & -0.01        \\
IC4756  & 1.26                       & -0.12        \\
NGC2345 & 1.37                       & -0.22        \\
NGC2423 & 1.46                       & -0.03        \\
NGC3114 & 1.46                       & -0.10        \\
NGC3532 & 1.40                       & -0.08        \\
NGC3680 & 1.26                       & -0.15        \\
NGC4349 & 1.26                       & -0.11        \\
NGC5822 & 1.46                       & -0.10        \\
NGC6705 & 1.54$^{(1)}$               & 0.07         \\ \bottomrule
\end{tabular}
\tablefoot{The note (1) indicates A(Li$)_{\text{max}}$ was derived in \cite{Randich2020,Romano2021}. The remainder of these limits and metallicities were derived in \cite{MariaLi}.}
\label{table:li_limits}
\end{table}

\begin{table*}[h!]
\centering
\caption{Lithium-rich and `Li-Rich-in-Context' giants present in our sample.}
\resizebox{\textwidth}{!}{
\begin{tabular}{@{}cccccccccc@{}}
\toprule
Star          & A(Li)$_{\textrm{NLTE}}$ [dex] & Li-Rich Type & Evolutionary Stage & Mass [M$_\odot$]&$v\sin(i)$ [\kms]& \ce{^{12}C/^{13}C}& Possible Li Production Mechanism\\ \midrule
IC4651No6333 & 1.21& In Context &  luminosity bump& 1.78& 2.13&-& Extra-Mixing \\
IC4651No9791 & 1.28& In Context &luminosity bump &1.75 & 1.85&-& Extra-Mixing\\
 \midrule
IC4756No52 $^{(\bullet)}$   & 1.45  &   $>$A(Li)$_\text{max}$   & RGB tip &2.27&2.87 & -& Unknown\\\midrule
NGC2345No50 & 0.98& In Context &RGB/early AGB & 5.84 & 5.27&(>)26$^{(\xi)}$ & Unknown/HBB \\\midrule
NGC2423No3 & 1.38& In Context &RGB-Lum. bump? &2.04 & 2.19 &- & Unknown \\\midrule
NGC3114No170$^{(\sharp)}$   & 1.50   &  Canonical   & red clump &3.98&11.32 & ${(\geq)}$ 24 $^{(\sharp\sharp)}$& Planet Accretion/Tidal Interactions \\\midrule
NGC3532No19   & 1.44    &  $>$A(Li)$_\text{max}$   &red clump &3.22& 5.05 & 12$^{(\dag)}$ & Extra-Mixing$^{(\delta)}$\\
NGC3532No649$^{(\flat)}$  & 3.27  &   Canonical    & -&2.17& -& 10$^{(\ddag)}$& Extra-Mixing\\
NGC3532No670  & 1.47    &   $>$A(Li)$_\text{max}$  & RGB tip &3.05& 4.64& 20$^{(\dag)}$ & Unknown \\\midrule
NGC3680No13 & 1.25& In Context & luminosity bump& 1.66& 0.28& 16 $^{(\zeta)}$ &Extra-Mixing  \\
NGC3680No53 & 1.21& In Context & luminosity bump& 1.65& 0.12& 10 $^{(\zeta)}$& Extra-Mixing \\\midrule
NGC4349No9  & 1.26   &   $=$A(Li)$_\text{max}$   & red clump&3.00&9.58 & 16$^{(\ast)}$ & Planet Accretion/Tidal Interactions\\
NGC4349No127  & 1.37   &  $>$A(Li)$_\text{max}$    & RGB tip&3.01&4.81 & 18$^{(\ast)}$ &  Unknown\\\midrule
NGC5822No102 & 1.34& In Context & RGB base/red clump & 2.02 &5.83 &- & Unknown\\
NGC5822No240 & 1.40& In Context &RGB tip &2.07 &2.83 & 17$^{\dag}$& Extra-Mixing \\\midrule
NGC6705No1101 & 1.53   &   Canonical   & RGB base &3.67&11.24 & -& Li-rich Formation Environment\\
NGC6705No1117 & 1.57   &  Canonical     & red clump&3.52&9.44 & -& Li-rich Formation Environment\\
NGC6705No1248 & 1.52   &   Canonical    & red clump&3.46&8.43 & -& Li-rich Formation Environment\\
NGC6705No1256 & 1.59   &   Canonical    & RGB &3.15& 4.47&-& Li-rich Formation Environment \\
NGC6705No1364 & 1.53   &  Canonical     & red clump &3.67&12.24 & -& Li-rich Formation Environment \\
NGC6705No1658 & 1.51    &  Canonical    & red clump &3.91&7.29 & -& Li-rich Formation Environment \\
NGC6705No411  & 1.57     &  Canonical   & RGB &3.19&6.47 &- & Li-rich Formation Environment \\
NGC6705No660  & 1.53     &  Canonical   & red clump&3.58&5.74 &- & Li-rich Formation Environment \\
NGC6705No669  & 1.51    &  Canonical   & red clump&3.45&8.93 &- & Li-rich Formation Environment \\
NGC6705No779  & 1.57    &  Canonical    & RGB &3.40&5.57 & -& Li-rich Formation Environment  \\
NGC6705No816  & 1.53     &  Canonical   & red clump&3.96&8.77 & -& Li-rich Formation Environment  \\
NGC6705No916  & 1.53    &   Canonical   & red clump&3.75&9.70 & -& Li-rich Formation Environment  \\
NGC6705No963  & 1.54    &   Canonical   & red clump&3.76&6.71 & -& Li-rich Formation Environment  \\\bottomrule
\end{tabular}}

\tablefoot{We classify stars as lithium-rich according to the canonical limit (1.5 [dex]) and the A(Li) limits listed in Table \ref{table:li_limits}, $>$A(Li)$_\text{max}$. We also list `Li-Rich-in-Context' stars. Stellar parameters and abundances are as derived in \citet{MariaLi} or otherwise specifically stated.}

\tablebib{($\delta$) This work; ($\sharp$) Possible non-member: \cite{Santrich2013}. ($\sharp\sharp$) As derived in \cite{Santrich2013}. ($\dag$) As derived in \cite{Smiljanic2009}. ($\ddag$) As derived in \cite{Luck1994}. ($\flat$) Non-member: \cite{Mermilliod2008}. ($\bullet$) Possible non-member: \cite{Herzog1975,Frinchaboy2008,Bagdonas2018}. ($\ast$) As derived in \cite{Holanda2022}. ($\xi$) As derived in \cite{Holanda2019}. ($\zeta$) As derived in \cite{PenaSuarez2018}.}
\label{table:lirich}
\end{table*}

\section{Methods}
\label{methods}

To probe if the high A(Li) that we observe in red giant stars is connected to their stellar activity, we {explored} two methods. First, we studied how Li abundances relate to the following stellar activity indicators:
\begin{enumerate}
    \item the bisector inverse slope (BIS) of the cross correlation function;
    \item the full width at half maximum (FWHM) of the cross correlation function;
    \item the H$_\alpha$ index with both a 0.6 \AA   \hspace{0.5mm} and 1.6 \AA \hspace{0.5mm} central bandpass (\has, \hal);
    \item and the effective temperature (T$_{\textrm{eff}}$).
 
\end{enumerate}
In particular, we looked for relations between A(Li) and the {standard deviation (STD) and peak-to-peak amplitude (PTP)} of all stellar activity indicators as well as the mean and median of \ha . Only statistically significant models at a 99\% confidence level (with p-values of the F-test $<0.01$) are discussed.

Secondly, we investigated if a possible connection between A(Li) and stellar activity manifests in RV variability. So, we sought correlations between stellar activity indicators and RV measurements {(see Figure \ref{fig:exmpl_corrs})}. These were computed according to the Pearson correlation coefficient, $\rho_w$, which is weighted here by the uncertainty in the measurements. The p-values for each correlation are also provided to attest to the statistical significance of the correlations. A p-value bellow 0.05 suggests that the correlation is significantly different from zero at a 95\% confidence level. In addition, periodic temporal variations of the RV measurements of these stars and the stellar activity indicators were analysed by drawing generalised Lomb-Scargle Periodograms (GLSPs \citep{Zechmeister09}) in order to evaluate if they connect to the presence of Li. 

When searching for planets, it is important to devise methods to distinguish planetary signals from temporal variability that may arise from the phenomena appearing at the surface of a star as a consequence of stellar magnetic activity. The BIS and FWHM values of each observation are provided by the HARPS data reduction pipeline as part of the process to calculate the RV of a given spectrum. Correlations found between RV and respective BIS and FWHM measurements indicate that periodical variability noticed in RV signals may originate from stellar activity or stellar surface inhomogeneities rather than planets. Variability observed in BIS and FWHM over time with approximately the same period as that observed in RV can also further corroborate this. These types of studies have often been conducted to probe whether or not a planet really exists in the orbit of a certain star (e.g. \citealt{queloz2001,ElisaPlanets}).

In studying how stellar activity impacts an RV signal, it is common to analyse spectral lines that serve as an activity proxy due to their well-known correlations with the existence of active regions on the surface of a star. One such spectral line is the H$_\alpha$ line. Here, we determined H$_\alpha$ index values from spectra using ACTIN 2\footnote{\url{https://github.com/gomesdasilva/ACTIN2}} \citep{Da_Silva_2018,2021JGS}. This tool to calculate stellar activity indices provides two measurements for the H$_\alpha$ index for two different central bandpasses: 0.6 \AA \hspace{0.5mm} and 1.6 \AA\ {(see Figure \ref{fig:haband})}. Since it has been mostly used for dwarf stars \citep{daSilva2022}, there is a lack of an extensive study for stars such as the ones in our sample, and it is not clear which central bandpass is optimal for the study of these stars. Recently, in analysing the strong RV signals of some of the stars in this sample \citep{Elisa2023}, we found that \hal\ seems to be a more reliable activity indicator for these evolved stars. But this conclusion was based on a small sample. As such, and since the two bandpasses probe the stellar atmosphere at different layers that are sensitive to different phenomena, we used both bandpasses in our analysis.  

We further extended our search for correlations and periodical variations to the \teff. Manifestations of magnetic activity, much like spots and plages are usually accompanied by temperature variations. Spots are cooler than their surrounding photosphere, while plages are hotter than their surrounding chromosphere. During periods of intense magnetic activity, it is possible that {if the \teff\ is affected}, the variations observed could correlate with the activity-induced variations \citep[e.g.][]{yana-galarza19}. {We used the stellar spectral analysis package FASMA\footnote{\url{https://github.com/MariaTsantaki/FASMA-synthesis}} \citep{FASMA} to determine the \teff\ from individual stellar spectra. FASMA relies on the radiative transfer code MOOG\footnote{\url{https://www.as.utexas.edu/~chris/moog.html}} \citep{sneden1973} to model the atmosphere and create a synthetic spectrum for a set of input parameters of a given star, which include surface gravity, $\log(g)$; iron metallicity, [Fe/H]; macro-turbulence velocity $v_{\textrm{mac}}$ ; and projected rotational velocity $v\sin(i)$. The synthetic spectrum was then statistically compared to the observed spectrum of the star, and the parameters were iteratively adjusted using a non-linear least-squares algorithm in order to find the set that best fits the observations. To capture reliable variations in \teff, when computing it, we fixed all other input stellar parameters and left \teff\ to be computed freely for each individual stellar spectra. Stellar parameters for the stars in our sample have been derived in \citet{MariaLi} also by using FASMA.}

\section{Results and discussion}
\label{discres}

As presented in Table \ref{table:lirich} and in accordance with \cite{Mermilliod2008}, the star {NGC3532No649} is not a member of cluster NGC3532, making the calculation of its mass and evolutionary stage uncertain. For this reason, it is not included in the following discussion despite it being Li-rich.

\begin{figure}[h!]
\centering
    \includegraphics[width=\columnwidth]{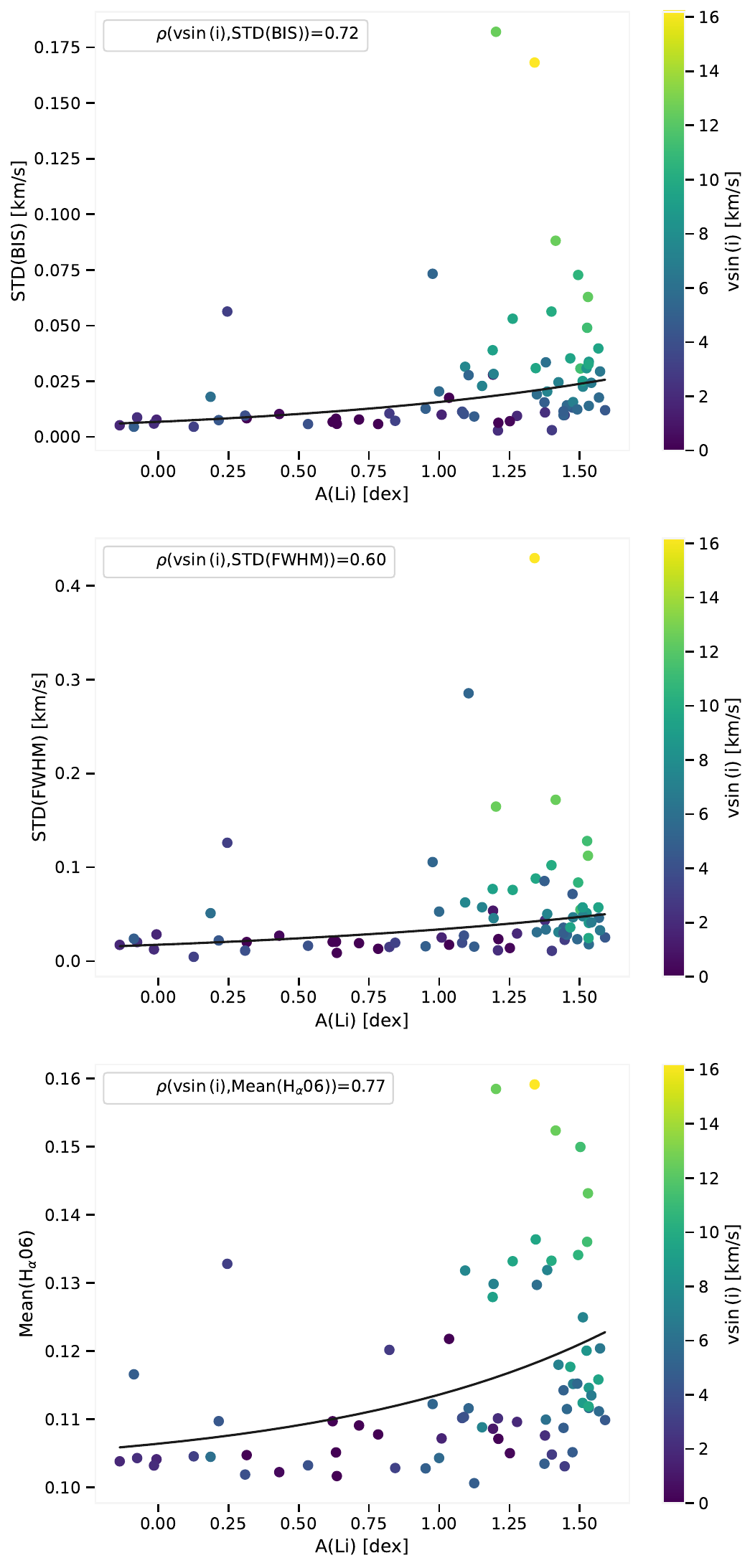}
    \caption{Exponential relation found between the STD of BIS data for each star and A(Li) [top panel], the STD of FWHM data for each star and A(Li) [middle panel], and the mean of \has\ data for each star and A(Li) [bottom panel]. The data points are coloured according to the $v\sin(i)$ of each star. {The black lines are the lines of best fit for which the F-test statistics and p-values are listed in Table \ref{exp}.} In these plots we also show the Pearson correlation coefficients ($\rho$) between the activity proxies used and $v\sin(i)$; all of them are very strong ($>0.6$).}
    \label{LiVsSTD}
\end{figure}

\subsection{The relationship between lithium abundance and stellar activity in red giants}

We started by analysing the dispersion (STD and PTP) of BIS, FWHM, T$_{\text{eff}}$, and \ha\ in order to probe for any potential connections between activity variability and A(Li). We also tested if the mean and median of H$_\alpha$, which are indicative of a star's activity level, are related to A(Li). For the computation of these relations, we only used stars with A(Li)$>-0.5$, as this is the limit of FASMA's minimisation algorithm in the determination of A(Li) \citep{MariaLi}. Using these limit values would be a source of bias. We find that the STD and PTP of BIS and FWHM increase exponentially with A(Li), and the mean and median values of \has \hspace{0.5mm} also increase exponentially with A(Li). In particular, the vast majority of stars with a high \has\ index are Li enriched. On the other hand, we did not find A(Li) to be related to \hal. These results suggest that the activity levels of the stars in our sample seem to increase with increasing A(Li). These results are in agreement with \cite{sneden22}, who found that Li-rich giants are more chromospherically active in comparison to their Li-poor counterparts, as they are more likely to show strong He I$\lambda$10830 absorptions. We also find that $v\sin(i)$ correlates well with the dispersion of BIS and FHWM as well as the mean and median of the \has\ index. This is indicative that more magnetically active stars also tend to rotate faster. {\cite{Goncalves20} also arrived at a similar conclusion.} {
The best-fit relations are listed in Table \ref{exp}. Plots of the relations found for the STD of BIS and FWHM as well as the mean of \has\ are shown in Figure \ref{LiVsSTD}. The plots for the PTP of BIS and FWHM and the median of \has\ look similar to the ones shown in Figure \ref{LiVsSTD} and are presented in Appendix \ref{SSR}. Also in Appendix \ref{SSR} are plots of the mean and median \hal\ as a function of A(Li), for which no clear relation was found.} In addition to the results of the F-test, in Table \ref{exp} we also provide R$^2$ values, with better fits having R$^2$ closer to one. The R$^2$ values for the relations we found are in general low, as there is significant scatter around the lines of the best fit.

\begin{table}[h!]
\centering
\caption{Best-fit relations of the descriptive statistics of the stellar activity indicators and $v\sin(i)$ as a function of A(Li).}
\resizebox{\columnwidth}{!}{
\begin{tabular}{@{}cccc@{}}
\toprule
Formula                                      & F-Statistic & p(F-Statistic) & R$^2$ \\ \midrule
STD(BIS)=$0.0067e^{0.84\text{A(Li)}}$               & 20.62       & 2.2$\times 10^{-5}$         & 0.22  \\
STD(FWHM)=$0.017e^{0.66\text{A(Li)}}$               & 13.59       & 0.0004         & 0.16  \\
PTP(BIS)=$0.024e^{0.84\text{A(Li)}}$                & 18.66      & 4.93$\times 10^{-5}$       & 0.21  \\
PTP(FWHM)=$0.060e^{0.66\text{A(Li)}}$               & 11.73       & 0.001         & 0.14  \\
Mean$_{H\alpha06}$=$0.004e^{\text{A(Li)}}+0.1$ & 11.55       & 0.001          & 0.14  \\
Median$_{H\alpha06}$=$0.004e^{\text{A(Li)}}+0.1$            & 12.12       & 0.0009          & 0.14  \\ 
$v\sin(i)$=$1.69e^{\text{A(Li)}}-0.2$        & 34.82       & 1.09$\times 10^{-7}$   & 0.33  \\
\bottomrule
\end{tabular}}
\tablefoot{Every relation is accompanied by the respective F-statistic and its p-value as well as R$^2$.} \label{exp}
\end{table}

\subsection{Manifestation of stellar activity in radial velocity signals and its connection to lithium enhancement}

\begin{figure}[h!]
    \centering
    \includegraphics[width=\columnwidth]{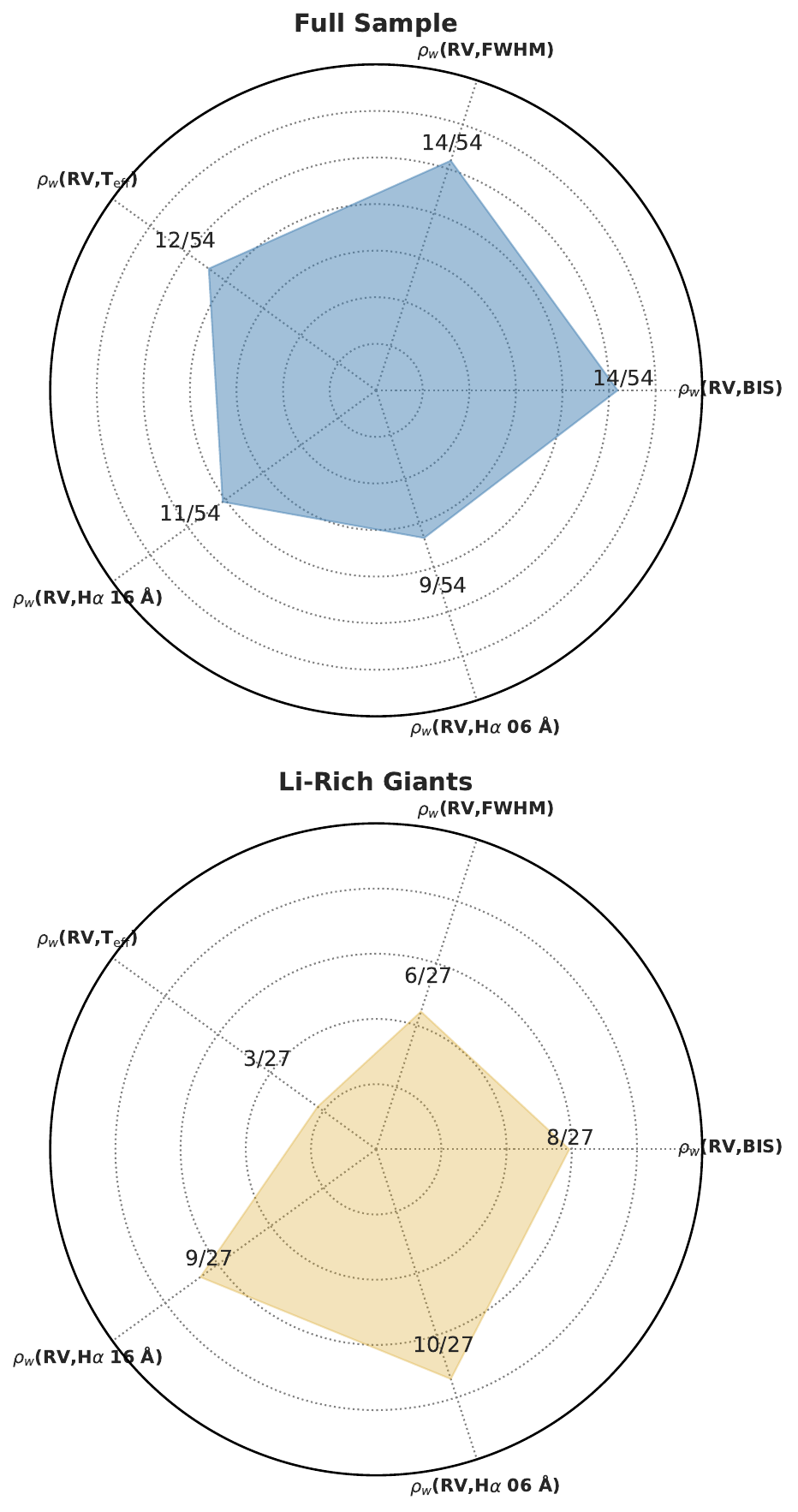}
    \caption{Spider charts showing the number of strong correlations found. The top chart shows how many strong correlations were found between RV and every stellar activity indicator in the full sample of red giant stars. The bottom chart shows how many strong correlations were found between RV and every stellar activity indicator for the stars of interest listed in Table \ref{table:lirich}.}
    \label{fig:corr_mtrx}
\end{figure}

To further investigate these findings, we also looked at how RV variability correlates with the stellar activity indicators and whether this is linked to A(Li). All the correlations discussed here are {shown in Figure \ref{fig:corr_mtrx} and listed in Appendix \ref{Corrs}}. From our analysis, 22/81 stars show a strong correlation ($>$0.4\footnote{Some of the analysed stars in our sample do not have a large number of data points. Because of this, we chose to include moderate correlations that have the potential to become strong with the addition of new data. For ease of nomenclature, we refer to all correlations greater than 0.4 as strong.}) between the RV {variability} and the BIS {variability}. The majority of the correlations were found to be negative. As stars age, their $v\sin(i)$ decreases, typically falling below 5 km/s by the end of the main-sequence \citep[e.g.][]{1996Medeiros,Tayar2015,2023Patton}. We noticed that 13/22 of the strong correlations found between RV and BIS appear for stars that are moderate and fast rotators ($v\sin(i) > 5\ \text{km/s}$). {This is likely a consequence of this diagnostic of line asymmetry losing sensitivity for slower rotators \citep[e.g.][]{2003Santos}}. There is, however, evidence in the literature (e.g. \citealt{Carlberg2012,sneden22,Sayeed2023,MariaLi}) that there is a connection between Li enrichment in giants and $v\sin(i)$. This is also suggested in our sample, as we find that $v\sin(i)$ increases exponentially with A(Li) ({shown in Appendix \ref{SSR} and Table \ref{exp}}). Out of the 27 Li-rich stars listed in Table \ref{table:lirich}, eight stars show strong correlations between RV and BIS.

In what concerns the FWHM, 20/81 stars show strong correlations with RV measurements. The correlations are preferentially positive. Six of the Li-rich stars in our sample show {strong} correlations between RV and FWHM. 

For the T$_{\textrm{eff}}$, we find that 15/81 stars show strong correlations with RV measurements. Correlations between the RV and \teff\ are mostly positive. Only three of the Li-rich stars show strong correlations between RV and T$_{\text{eff}}$. There is a possibility that this the case is because the individual spectra used to measure \teff\ values to look for variations may not have a signal-to-noise ratio high enough for this purpose.

As for \hal, 20/81 stars show strong correlations with RV, and they are mostly negative. Nine of the Li-rich stars show strong correlations between RV and \hal.

Last but not least, for \has, 19/81 stars show significant correlations with RV measurements, and they are preferentially positive. We find that ten of the Li-rich stars in our sample show strong correlations between RV and \has. 

\begin{figure}
    \centering
    \includegraphics[width=\columnwidth]{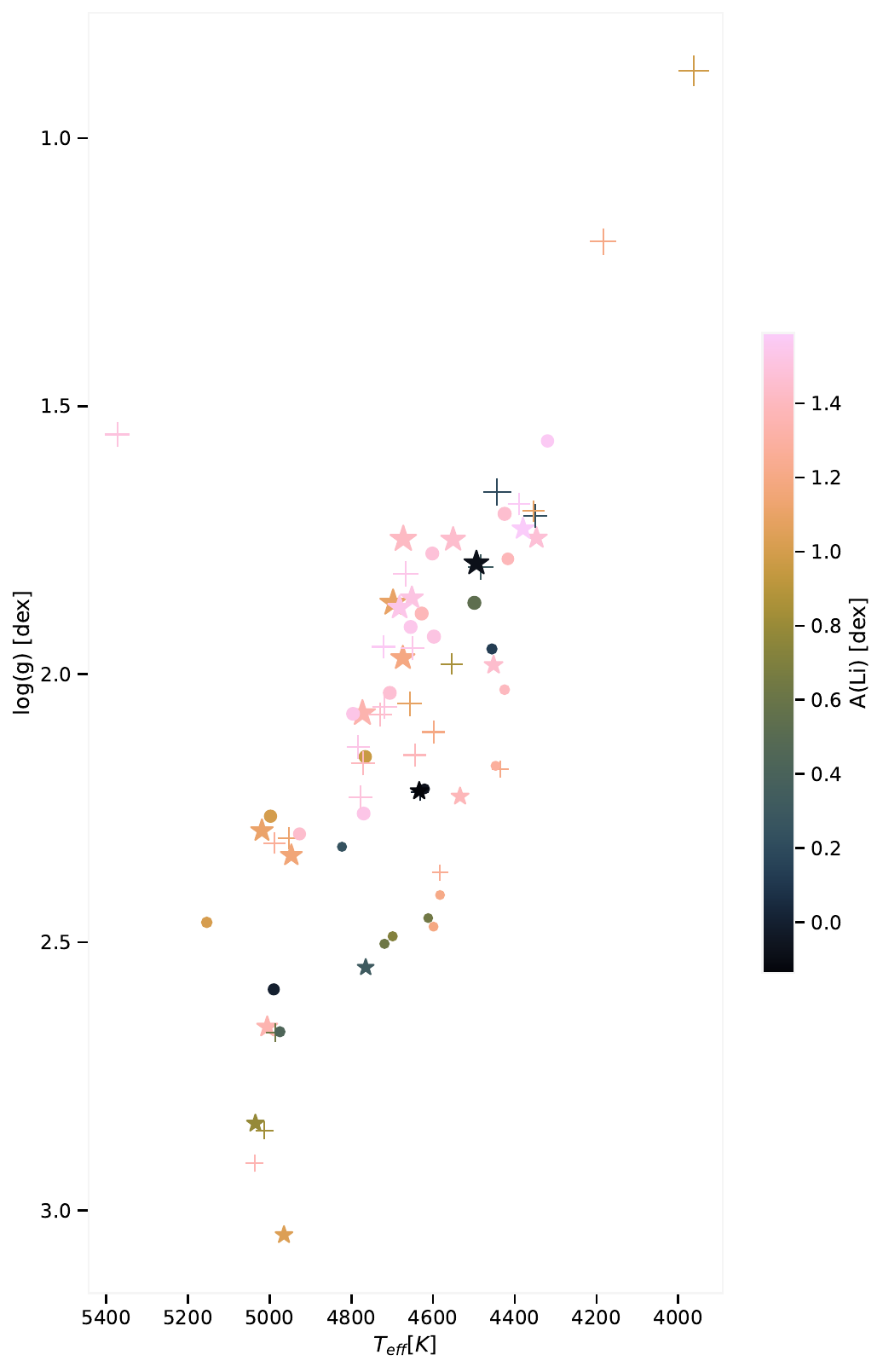}
    \caption{Hertzsprung–Russell diagram of all the stars in our sample. Here we use the $\log(g)$ and \teff\ derived in \cite{MariaLi} for each star. The stars are coloured according to A(Li) and have different sizes to represent their different masses. The different symbols indicate how many strong correlations were found in a particular star: "$\bullet$" indicates zero strong correlations, "$\star$" indicates one strong correlation, and "$+$" indicates two or more strong correlations.}
    \label{fig:HRDiam}
\end{figure}

Overall, correlations were found for stars in various evolutionary stages and of different mass ranges as well. {This is evidenced in the Hertzsprung–Russell (HR) diagram in Figure \ref{fig:HRDiam}, where different symbols indicate different numbers of strong correlations found per star and the data points are sized according to stellar mass. In general, all symbols display a range of sizes and are distributed across the HR diagram, revealing there is no preferred mass or age for correlations to be found.} A summary of the significant correlations discussed can be found in Tables \ref{summarypart1} and \ref{summarypart2}. Something to note from Table \ref{summarypart2} is that inhabitants of cluster NGC6705, in general, show higher amounts of Li than the rest of the stars in our sample. These stars are particular, as they also demonstrate unusually high amounts of $\alpha$-elements for their young age (e.g. \citealt{Casam2018}). \cite{Messina2010} found that the G-type stars of this cluster retain higher levels of activity up to the age of the cluster. We find that 19/28 of these stars show at least one significant correlation with one of the parameters we studied, {and 22/28 are fast rotators (see Table \ref{summarypart2})}. So, in general, the stars in this cluster are magnetically active and also rotate fast, which could be connected to the abnormally high levels of Li observed.

\begin{figure}[h!]
    \centering
    \includegraphics[width=\columnwidth]{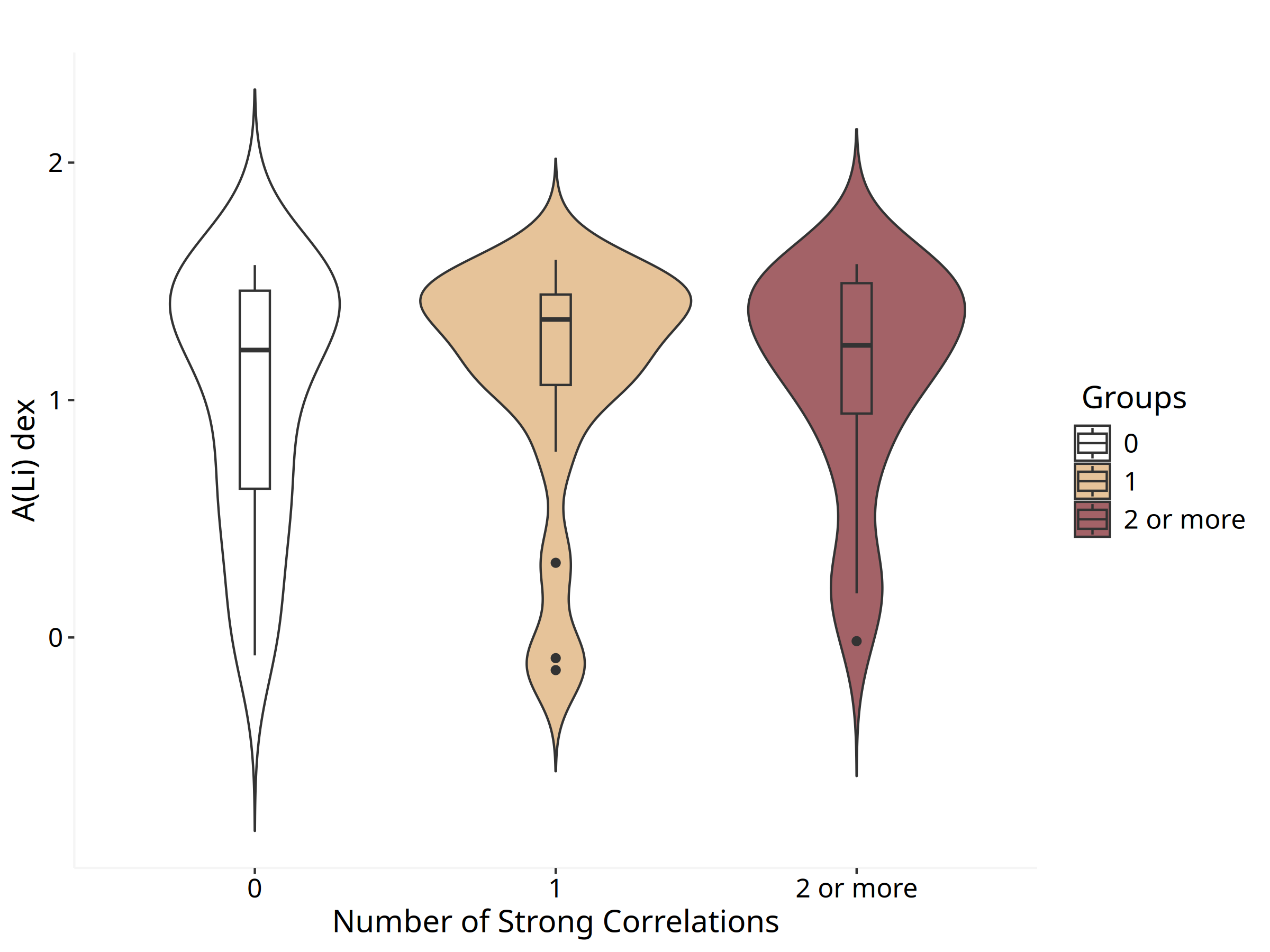}
    \caption{Violin plot of A(Li) as a function of the number of strong correlations observed per star. From left to right, each violin represents the distribution of the number of strong correlations found (0, 1, or $\geq 2$) relative to A(Li). Inside each violin is a box delimiting the interquartile range. The thick black line appearing inside each box is the median of the distribution. Each box also has a lower and upper whisker that spans from minimum to maximum A(Li) in each distribution. Points seen beyond the whiskers are outliers.}
    \label{fig:violin}
\end{figure}

In Figure \ref{fig:violin}, we show a violin plot of A(Li) as a function of the number of strong correlations observed per star. One may see that the density of the points increases towards higher values of A(Li), with an increasing number of correlations found. This is evidenced by the increasing width of the violins from left to right in high A(Li) regions and by the central tendency being slightly higher for stars that show one or more than two correlations between RV and stellar activity indicators. As such, strong correlations tend to appear more often for stars with higher amounts of Li in their atmospheres. Such a result also indicates that A(Li) seems to be connected with higher levels of stellar activity in our sample.

To complement the correlation analysis, GLSPs of RV and stellar activity indicators were also plotted for all stars. The GLSPs are often used when analysing whether variability observed in RV signals is caused by the existence of a companion or by stellar activity. This is because they make it possible to determine if there is periodic variability occurring at the stellar surface that is equivalent to the periodic variability observed in RV. It has been shown that magnetic phenomena appearing at the stellar surface can introduce variability in RV that is at a different phase relative to the variability observed in stellar activity indicator signals \citep[e.g.][]{queloz2001,2014Santos,ElisaPlanets}. As such, when looking for correlations between RV and activity proxies, we may obtain a weak correlation coefficient, but this does not necessarily mean that the RV variability does not originate from stellar activity. Thus, finding peaks at similar periods in RV and again in different stellar activity signals with GLSP is also an indicator that RV variability connects with stellar activity.

The results of the GLSP analysis are also presented in Tables \ref{summarypart1} and \ref{summarypart2}. In summary, stars NGC2345No50, NGC2423No3, NGC3532No670, NGC4349No127, and NGC6705No1101 show significant peaks ({above a false-alarm probability (FAP) level of 1\%}) in their RV GLSPs, which consistently appear around the same period for at least one of their stellar activity indicators. All relevant GLSPs are in Appendix \ref{LS} and \cite{Elisa2023}. All of these stars can be considered Li-rich (see Table \ref{table:lirich}), and most of them also present long-term and large amplitude RV variations that mimic the presence of planets but are of probable stellar origin \citep{Elisa2023}. These stars with periodic RV variations lie on the RGB or RGB tip (they are among the most luminous in their cluster) rather than in the red clump, as found in recent works for super-Li-rich stars \citep{Singh19,Kumar20}. Interestingly, \citet{Jorissen20}, studying the binarity fraction of Li-rich giants, found that some of the most luminous Li-rich giants show RV variability not caused by stellar companions, though no information about stellar activity is provided by the authors. Similarly, \citet{Goncalves20} found hints of RV variability in several Li-rich stars with a detected longitudinal magnetic field, but the quantity of data does not permit one to discern {the real origin of such variability}.

\section{Conclusions}
\label{conc}

In summary, based on the results from the HARPS data, we find that the variability of BIS and FWHM and the mean and median of \has \hspace{0.5mm} increase exponentially with A(Li). This leads us to conclude that the stars in our sample that are more magnetically active also show higher amounts of Li in their atmosphere. We also find that stars with higher amounts of Li tend to show more correlations between RV and stellar activity indicators. The majority of the stars, {19/27} (10/14 leaving aside the special cluster NGC6705), that can be considered Li-rich in this work either show correlations of RV with at least one of the stellar activity indicators or show periodic variability in at least one of the stellar activity indicator signals with a similar period in RV. From the GLSP analysis we may also conclude that almost all the stars with strong RV variations that mimic the presence of planets are Li-rich \cite[see also,][]{mena2016searching}. This is particularly important when searching for exoplanets orbiting evolved stars since our results show that detecting periodic RV variability in red giants with high A(Li) might be a signature of stellar activity instead of planets. It is interesting to note that \citet{Adamow18} reported a higher than expected fraction of planetary companions around the Li-rich stars in their sample, pointing to a connection between Li enrichment and planetary presence or engulfment. However, the orbits they found are quite eccentric, something not expected if the RV is of stellar origin, {and that characteristic supports the hypothesis of inner planet engulfment. The Li-rich stars with strong RV variability discussed here lie on the RGB or RGB tip. However, stars showing correlations between Li abundance and activity level are found all across the HR diagram. This might indicate that mild Li enrichment (since we do not have super Li-rich giants in our sample) connected to stellar activity can occur at different evolutionary stages, but a large enhancement of Li is more particular and can only occur at certain evolutionary stages, such as the He flash stage \citep{Singh19,Kumar20}. {Thus, it is still not clear whether the strong RV variability of mild Li-rich giants on the RGB is connected to their evolutionary stage and the possible super Li-rich phase that happens soon after (at the beginning of the He-core burning phase) or if it is a consequence of stellar activity}.

\begin{acknowledgements}
     We thank the anonymous referee for the insightful comments and suggestions. I.R. and E.D.M. acknowledge the support from Funda\c{c}\~ao para a Ci\^encia e a Tecnologia (FCT) through national funds and from FEDER through COMPETE2020 by the following grants: UIDB/04434/2020 \& UIDP/04434/2020. I.R. acknowledges the support from FCT through the grant 2022.03993.PTDC. E.D.M. acknowledges the support from FCT through Stimulus FCT contract 2021.01294.CEECIND. E.D.M and J.G.S. acknowledge the support from FCT through the grant 2022.04416.PTDC. This work made use of scientific colour maps \citep[\url{https://www.fabiocrameri.ch/colourmaps/}][]{crameri2020misuse} to mitigate visual distortion.
\end{acknowledgements}

\bibliography{Bibliography}

\begin{thebibliography}{76}
\expandafter\ifx\csname natexlab\endcsname\relax\def\natexlab#1{#1}\fi

\bibitem[{{Adam{\'o}w} {et~al.}(2018){Adam{\'o}w}, {Niedzielski}, {Kowalik}, {Villaver}, {Wolszczan}, {Maciejewski}, \& {Gromadzki}}]{Adamow18}
{Adam{\'o}w}, M., {Niedzielski}, A., {Kowalik}, K., {et~al.} 2018, \aap, 613, A47

\bibitem[{{Adam{\'o}w} {et~al.}(2012){Adam{\'o}w}, {Niedzielski}, {Villaver}, {Nowak}, \& {Wolszczan}}]{Adamow2012}
{Adam{\'o}w}, M., {Niedzielski}, A., {Villaver}, E., {Nowak}, G., \& {Wolszczan}, A. 2012, \apjl, 754, L15

\bibitem[{{Adam{\'o}w} {et~al.}(2014){Adam{\'o}w}, {Niedzielski}, {Villaver}, {Wolszczan}, \& {Nowak}}]{Adamow2014}
{Adam{\'o}w}, M., {Niedzielski}, A., {Villaver}, E., {Wolszczan}, A., \& {Nowak}, G. 2014, \aap, 569, A55

\bibitem[{{Aguilera-G{\'o}mez} {et~al.}(2016){Aguilera-G{\'o}mez}, {Chanam{\'e}}, {Pinsonneault}, \& {Carlberg}}]{Aguilera2016}
{Aguilera-G{\'o}mez}, C., {Chanam{\'e}}, J., {Pinsonneault}, M.~H., \& {Carlberg}, J.~K. 2016, \apj, 829, 127

\bibitem[{{Aguilera-G{\'o}mez} {et~al.}(2023){Aguilera-G{\'o}mez}, {Jones}, \& {Chanam{\'e}}}]{2023Aguilera}
{Aguilera-G{\'o}mez}, C., {Jones}, M.~I., \& {Chanam{\'e}}, J. 2023, \aap, 670, A73

\bibitem[{{Bagdonas} {et~al.}(2018){Bagdonas}, {Drazdauskas}, {Tautvai{\v{s}}ien{\.{e}}}, {Smiljanic}, \& {Chorniy}}]{Bagdonas2018}
{Bagdonas}, V., {Drazdauskas}, A., {Tautvai{\v{s}}ien{\.{e}}}, G., {Smiljanic}, R., \& {Chorniy}, Y. 2018, \aap, 615, A165

\bibitem[{{Brown} {et~al.}(1989){Brown}, {Sneden}, {Lambert}, \& {Dutchover}}]{Brown1989}
{Brown}, J.~A., {Sneden}, C., {Lambert}, D.~L., \& {Dutchover}, Edward, J. 1989, \apjs, 71, 293

\bibitem[{{Cameron} \& {Fowler}(1971)}]{CameronFowler1971}
{Cameron}, A.~G.~W. \& {Fowler}, W.~A. 1971, \apj, 164, 111

\bibitem[{{Cantiello} \& {Langer}(2010)}]{Cantiello2010}
{Cantiello}, M. \& {Langer}, N. 2010, \aap, 521, A9

\bibitem[{{Carlberg} {et~al.}(2012){Carlberg}, {Cunha}, {Smith}, \& {Majewski}}]{Carlberg2012}
{Carlberg}, J.~K., {Cunha}, K., {Smith}, V.~V., \& {Majewski}, S.~R. 2012, \apj, 757, 109

\bibitem[{{Casamiquela} {et~al.}(2018){Casamiquela}, {Carrera}, {Balaguer-N{\'u}{\~n}ez}, {Jordi}, {Chiappini}, {Anders}, {Antoja}, {Miret-Roig}, {Romero-Gomez}, {Blanco-Cuaresma}, {Pancino}, {Aguado}, {del Pino}, {Diaz-Perez}, \& {Gallart}}]{Casam2018}
{Casamiquela}, L., {Carrera}, R., {Balaguer-N{\'u}{\~n}ez}, L., {et~al.} 2018, \aap, 610, A66

\bibitem[{{Casey} {et~al.}(2016){Casey}, {Ruchti}, {Masseron}, {Randich}, {Gilmore}, {Lind}, {Kennedy}, {Koposov}, {Hourihane}, {Franciosini}, {Lewis}, {Magrini}, {Morbidelli}, {Sacco}, {Worley}, {Feltzing}, {Jeffries}, {Vallenari}, {Bensby}, {Bragaglia}, {Flaccomio}, {Francois}, {Korn}, {Lanzafame}, {Pancino}, {Recio-Blanco}, {Smiljanic}, {Carraro}, {Costado}, {Damiani}, {Donati}, {Frasca}, {Jofr{\'e}}, {Lardo}, {de Laverny}, {Monaco}, {Prisinzano}, {Sbordone}, {Sousa}, {Tautvai{\v{s}}ien{\.{e}}}, {Zaggia}, {Zwitter}, {Delgado Mena}, {Chorniy}, {Martell}, {Silva Aguirre}, {Miglio}, {Chiappini}, {Montalban}, {Morel}, \& {Valentini}}]{2016Casey}
{Casey}, A.~R., {Ruchti}, G., {Masseron}, T., {et~al.} 2016, \mnras, 461, 3336

\bibitem[{{Charbonnel}(1994)}]{1994Charbonnel}
{Charbonnel}, C. 1994, \aap, 282, 811

\bibitem[{{Charbonnel}(1995)}]{Charbonnel1995}
{Charbonnel}, C. 1995, \apjl, 453, L41

\bibitem[{{Charbonnel} \& {Balachandran}(2000)}]{charbonnel2000nature}
{Charbonnel}, C. \& {Balachandran}, S.~C. 2000, \aap, 359, 563

\bibitem[{Crameri {et~al.}(2020)Crameri, Shephard, \& Heron}]{crameri2020misuse}
Crameri, F., Shephard, G.~E., \& Heron, P.~J. 2020, Nature Communications, 11, 5444

\bibitem[{{de Medeiros} {et~al.}(1996){de Medeiros}, {Melo}, \& {Mayor}}]{1996Medeiros}
{de Medeiros}, J.~R., {Melo}, C.~H.~F., \& {Mayor}, M. 1996, \aap, 309, 465

\bibitem[{{Deepak} \& {Reddy}(2019)}]{2019Deepak}
{Deepak} \& {Reddy}, B.~E. 2019, \mnras, 484, 2000

\bibitem[{{Delgado Mena} {et~al.}(2023){Delgado Mena}, {Gomes da Silva}, {Faria}, {Santos}, {Martins}, {Tsantaki}, {Mortier}, {Sousa}, \& {Lovis}}]{Elisa2023}
{Delgado Mena}, E., {Gomes da Silva}, J., {Faria}, J.~P., {et~al.} 2023, \aap, 679, A94

\bibitem[{Delgado~Mena {et~al.}(2018)Delgado~Mena, Lovis, Santos, Gomes~da Silva, Mortier, Tsantaki, Sousa, Figueira, Cunha, Campante, Adibekyan, Faria, \& Montalto}]{ElisaPlanets}
Delgado~Mena, E., Lovis, C., Santos, N.~C., {et~al.} 2018, A\&A, 619, A2

\bibitem[{{Delgado Mena} {et~al.}(2016){Delgado Mena}, {Tsantaki}, {Sousa}, {Kunitomo}, {Adibekyan}, {Zaworska}, {Santos}, {Israelian}, \& {Lovis}}]{mena2016searching}
{Delgado Mena}, E., {Tsantaki}, M., {Sousa}, S.~G., {et~al.} 2016, \aap, 587, A66

\bibitem[{{Denissenkov} \& {Herwig}(2004)}]{2004Denissenkov}
{Denissenkov}, P.~A. \& {Herwig}, F. 2004, \apj, 612, 1081

\bibitem[{{Eggleton} {et~al.}(2008){Eggleton}, {Dearborn}, \& {Lattanzio}}]{Egg2008}
{Eggleton}, P.~P., {Dearborn}, D. S.~P., \& {Lattanzio}, J.~C. 2008, \apj, 677, 581

\bibitem[{{Favata} {et~al.}(1996){Favata}, {Micela}, \& {Sciortino}}]{favata1996lithium}
{Favata}, F., {Micela}, G., \& {Sciortino}, S. 1996, \aap, 311, 951

\bibitem[{Fekel \& Balachandran(1994)}]{fekel1994lithium}
Fekel, F.~C. \& Balachandran, S. 1994, in Cool Stars, Stellar Systems, and the Sun, Vol.~64, 279

\bibitem[{{Frinchaboy} \& {Majewski}(2008)}]{Frinchaboy2008}
{Frinchaboy}, P.~M. \& {Majewski}, S.~R. 2008, \aj, 136, 118

\bibitem[{{Gao} {et~al.}(2019){Gao}, {Shi}, {Yan}, {Yan}, {Xiang}, {Zhou}, {Li}, \& {Zhao}}]{Gao2019}
{Gao}, Q., {Shi}, J.-R., {Yan}, H.-L., {et~al.} 2019, \apjs, 245, 33

\bibitem[{{Gomes da Silva} {et~al.}(2022){Gomes da Silva}, {Bensabat}, {Monteiro}, \& {Santos}}]{daSilva2022}
{Gomes da Silva}, J., {Bensabat}, A., {Monteiro}, T., \& {Santos}, N.~C. 2022, \aap, 668, A174

\bibitem[{{Gomes da Silva} {et~al.}(2018){Gomes da Silva}, Figueira, Santos, \& Faria}]{Da_Silva_2018}
{Gomes da Silva}, J., Figueira, P., Santos, N., \& Faria, J. 2018, Journal of Open Source Software, 3, 667

\bibitem[{{Gomes da Silva} {et~al.}(2021){Gomes da Silva}, {Santos}, {Adibekyan}, {Sousa}, {Campante}, {Figueira}, {Bossini}, {Delgado-Mena}, {Monteiro}, {de Laverny}, {Recio-Blanco}, \& {Lovis}}]{2021JGS}
{Gomes da Silva}, J., {Santos}, N.~C., {Adibekyan}, V., {et~al.} 2021, \aap, 646, A77

\bibitem[{{Gon{\c{c}}alves} {et~al.}(2020){Gon{\c{c}}alves}, {da Costa}, {de Almeida}, {Castro}, \& {do Nascimento}}]{Goncalves20}
{Gon{\c{c}}alves}, B.~F.~O., {da Costa}, J.~S., {de Almeida}, L., {Castro}, M., \& {do Nascimento}, J.~D., J. 2020, \mnras, 498, 2295

\bibitem[{{Herzog} {et~al.}(1975){Herzog}, {Sanders}, \& {Seggewiss}}]{Herzog1975}
{Herzog}, A.~D., {Sanders}, W.~L., \& {Seggewiss}, W. 1975, \aaps, 19, 211

\bibitem[{{Holanda} {et~al.}(2019){Holanda}, {Pereira}, \& {Drake}}]{Holanda2019}
{Holanda}, N., {Pereira}, C.~B., \& {Drake}, N.~A. 2019, \mnras, 482, 5275

\bibitem[{Holanda {et~al.}(2022)Holanda, Ramos, Peña Suárez, Martinez, \& Pereira}]{Holanda2022}
Holanda, N., Ramos, A.~A., Peña Suárez, V.~J., Martinez, C.~F., \& Pereira, C.~B. 2022, Monthly Notices of the Royal Astronomical Society, 516, 4484

\bibitem[{{Iben}(1967{\natexlab{a}})}]{1967bIben}
{Iben}, Icko, J. 1967{\natexlab{a}}, \apj, 147, 650

\bibitem[{{Iben}(1967{\natexlab{b}})}]{1967aIben}
{Iben}, Icko, J. 1967{\natexlab{b}}, \apj, 147, 624

\bibitem[{{Jorissen} {et~al.}(2020){Jorissen}, {Van Winckel}, {Siess}, {Escorza}, {Pourbaix}, \& {Van Eck}}]{Jorissen20}
{Jorissen}, A., {Van Winckel}, H., {Siess}, L., {et~al.} 2020, \aap, 639, A7

\bibitem[{{Katime Santrich} {et~al.}(2013){Katime Santrich}, {Pereira}, \& {Drake}}]{Santrich2013}
{Katime Santrich}, O.~J., {Pereira}, C.~B., \& {Drake}, N.~A. 2013, \aap, 554, A2

\bibitem[{{Kirby} {et~al.}(2016){Kirby}, {Guhathakurta}, {Zhang}, {Hong}, {Guo}, {Guo}, {Cohen}, \& {Cunha}}]{Kirby_2016}
{Kirby}, E.~N., {Guhathakurta}, P., {Zhang}, A.~J., {et~al.} 2016, \apj, 819, 135

\bibitem[{{Kowkabany} {et~al.}(2022){Kowkabany}, {Ezzeddine}, {Charbonnel}, {Roederer}, {Li}, {Hackshaw}, {Beers}, {Frebel}, {Hansen}, {Holmbeck}, {Placco}, \& {Sakari}}]{Jeremy2022}
{Kowkabany}, J., {Ezzeddine}, R., {Charbonnel}, C., {et~al.} 2022, arXiv e-prints, arXiv:2209.02184

\bibitem[{{Kumar} {et~al.}(2020){Kumar}, {Reddy}, {Campbell}, {Maben}, {Zhao}, \& {Ting}}]{Kumar20}
{Kumar}, Y.~B., {Reddy}, B.~E., {Campbell}, S.~W., {et~al.} 2020, Nature Astronomy, 4, 1059

\bibitem[{{Kumar} {et~al.}(2011){Kumar}, {Reddy}, \& {Lambert}}]{2011Kumar}
{Kumar}, Y.~B., {Reddy}, B.~E., \& {Lambert}, D.~L. 2011, \apjl, 730, L12

\bibitem[{{Lagarde} {et~al.}(2024){Lagarde}, {Minkevi{\v{c}}i{\={u}}t{\.{e}}}, {Drazdauskas}, {Tautvai{\v{s}}ien{\.{e}}}, {Charbonnel}, {Reyl{\'e}}, {Miglio}, {Kushwahaa}, \& {Bale}}]{2024A&Lagarde}
{Lagarde}, N., {Minkevi{\v{c}}i{\={u}}t{\.{e}}}, R., {Drazdauskas}, A., {et~al.} 2024, \aap, 684, A70

\bibitem[{{Liu} {et~al.}(2014){Liu}, {Tan}, {Wang}, {Zhao}, {Sato}, {Takeda}, \& {Li}}]{2014Liu}
{Liu}, Y.~J., {Tan}, K.~F., {Wang}, L., {et~al.} 2014, \apj, 785, 94

\bibitem[{{Lovis} \& {Mayor}(2007)}]{LovisMayor2007}
{Lovis}, C. \& {Mayor}, M. 2007, \aap, 472, 657

\bibitem[{{Luck}(1994)}]{Luck1994}
{Luck}, R.~E. 1994, \apjs, 91, 309

\bibitem[{{Lyubimkov}(2016)}]{lyubimkov2016lithium}
{Lyubimkov}, L.~S. 2016, Astrophysics, 59, 411

\bibitem[{{Martell} {et~al.}(2021){Martell}, {Simpson}, {Balasubramaniam}, {Buder}, {Sharma}, {Hon}, {Stello}, {Ting}, {Asplund}, {Bland-Hawthorn}, {De Silva}, {Freeman}, {Hayden}, {Kos}, {Lewis}, {Lind}, {Zucker}, {Zwitter}, {Campbell}, {{\v{C}}otar}, {Horner}, {Montet}, \& {Wittenmyer}}]{2021Martell}
{Martell}, S.~L., {Simpson}, J.~D., {Balasubramaniam}, A.~G., {et~al.} 2021, \mnras, 505, 5340

\bibitem[{{Mayor} {et~al.}(2003){Mayor}, {Pepe}, {Queloz}, {Bouchy}, {Rupprecht}, {Lo Curto}, {Avila}, {Benz}, {Bertaux}, {Bonfils}, {Dall}, {Dekker}, {Delabre}, {Eckert}, {Fleury}, {Gilliotte}, {Gojak}, {Guzman}, {Kohler}, {Lizon}, {Longinotti}, {Lovis}, {Megevand}, {Pasquini}, {Reyes}, {Sivan}, {Sosnowska}, {Soto}, {Udry}, {van Kesteren}, {Weber}, \& {Weilenmann}}]{Mayor2003}
{Mayor}, M., {Pepe}, F., {Queloz}, D., {et~al.} 2003, The Messenger, 114, 20

\bibitem[{{Mermilliod} {et~al.}(2008){Mermilliod}, {Mayor}, \& {Udry}}]{Mermilliod2008}
{Mermilliod}, J.~C., {Mayor}, M., \& {Udry}, S. 2008, \aap, 485, 303

\bibitem[{{Messina} {et~al.}(2010){Messina}, {Parihar}, {Koo}, {Kim}, {Rey}, \& {Lee}}]{Messina2010}
{Messina}, S., {Parihar}, P., {Koo}, J.-R., {et~al.} 2010, A\&A, 513, A29

\bibitem[{{Pabst} {et~al.}(2020){Pabst}, {Goicoechea}, {Teyssier}, {Bern{\'e}}, {Higgins}, {Chambers}, {Kabanovic}, {G{\"u}sten}, {Stutzki}, \& {Tielens}}]{2020Pabst}
{Pabst}, C.~H.~M., {Goicoechea}, J.~R., {Teyssier}, D., {et~al.} 2020, \aap, 639, A2

\bibitem[{Palacios {et~al.}(2006)Palacios, Charbonnel, Talon, \& Siess}]{Palacios2004}
Palacios, A., Charbonnel, C., Talon, S., \& Siess, L. 2006, in Chemical Abundances and Mixing in Stars in the Milky Way and its Satellites, ed. S.~Randich \& L.~Pasquini (Berlin, Heidelberg: Springer Berlin Heidelberg), 304--305

\bibitem[{{Pallavicini} {et~al.}(1992){Pallavicini}, {Randich}, \& {Giampapa}}]{pallavicini1992lithium}
{Pallavicini}, R., {Randich}, S., \& {Giampapa}, M.~S. 1992, \aap, 253, 185

\bibitem[{{Patton} {et~al.}(2024){Patton}, {Pinsonneault}, {Cao}, {Vrard}, {Mathur}, {Garc{\'\i}a}, {Tayar}, {Daher}, \& {Beck}}]{2023Patton}
{Patton}, R.~A., {Pinsonneault}, M.~H., {Cao}, L., {et~al.} 2024, \mnras, 528, 3232

\bibitem[{{Pe{\~n}a Su{\'a}rez} {et~al.}(2018){Pe{\~n}a Su{\'a}rez}, {Sales Silva}, {Katime Santrich}, {Drake}, \& {Pereira}}]{PenaSuarez2018}
{Pe{\~n}a Su{\'a}rez}, V.~J., {Sales Silva}, J.~V., {Katime Santrich}, O.~J., {Drake}, N.~A., \& {Pereira}, C.~B. 2018, \apj, 854, 184

\bibitem[{Queloz {et~al.}(2001)Queloz, Henry, Sivan, Baliunas, Beuzit, Donahue, Mayor, Naef, Perrier, \& Udry}]{queloz2001}
Queloz, D., Henry, G.~W., Sivan, J.~P., {et~al.} 2001, A\&A, 379, 279

\bibitem[{{Queloz} {et~al.}(2000){Queloz}, {Mayor}, {Weber}, {Bl{\'e}cha}, {Burnet}, {Confino}, {Naef}, {Pepe}, {Santos}, \& {Udry}}]{Queloz2000}
{Queloz}, D., {Mayor}, M., {Weber}, L., {et~al.} 2000, \aap, 354, 99

\bibitem[{{Randich} {et~al.}(2020){Randich}, {Pasquini}, {Franciosini}, {Magrini}, {Jackson}, {Jeffries}, {d'Orazi}, {Romano}, {Sanna}, {Tautvai{\v{s}}ien{\.{e}}}, {Tsantaki}, {Wright}, {Gilmore}, {Bensby}, {Bragaglia}, {Pancino}, {Smiljanic}, {Bayo}, {Carraro}, {Gonneau}, {Hourihane}, {Morbidelli}, \& {Worley}}]{Randich2020}
{Randich}, S., {Pasquini}, L., {Franciosini}, E., {et~al.} 2020, \aap, 640, L1

\bibitem[{{Romano} {et~al.}(2021){Romano}, {Magrini}, {Randich}, {Casali}, {Bonifacio}, {Jeffries}, {Matteucci}, {Franciosini}, {Spina}, {Guiglion}, {Chiappini}, {Mucciarelli}, {Ventura}, {Grisoni}, {Bellazzini}, {Bensby}, {Bragaglia}, {de Laverny}, {Korn}, {Martell}, {Tautvai{\v{s}}ien{\.{e}}}, {Carraro}, {Gonneau}, {Jofr{\'e}}, {Pancino}, {Smiljanic}, {Vallenari}, {Fu}, {Guti{\'e}rrez Albarr{\'a}n}, {Jim{\'e}nez-Esteban}, {Montes}, {Damiani}, {Bergemann}, \& {Worley}}]{Romano2021}
{Romano}, D., {Magrini}, L., {Randich}, S., {et~al.} 2021, \aap, 653, A72

\bibitem[{{Santos} {et~al.}(2014){Santos}, {Mortier}, {Faria}, {Dumusque}, {Adibekyan}, {Delgado-Mena}, {Figueira}, {Benamati}, {Boisse}, {Cunha}, {Gomes da Silva}, {Lo Curto}, {Lovis}, {Martins}, {Mayor}, {Melo}, {Oshagh}, {Pepe}, {Queloz}, {Santerne}, {S{\'e}gransan}, {Sozzetti}, {Sousa}, \& {Udry}}]{2014Santos}
{Santos}, N.~C., {Mortier}, A., {Faria}, J.~P., {et~al.} 2014, \aap, 566, A35

\bibitem[{{Santos} {et~al.}(2003){Santos}, {Udry}, {Mayor}, {Naef}, {Pepe}, {Queloz}, {Burki}, {Cramer}, \& {Nicolet}}]{2003Santos}
{Santos}, N.~C., {Udry}, S., {Mayor}, M., {et~al.} 2003, \aap, 406, 373

\bibitem[{{Sayeed} {et~al.}(2024){Sayeed}, {Ness}, {Montet}, {Cantiello}, {Casey}, {Buder}, {Bedell}, {Breivik}, {Metzger}, {Martell}, \& {McGee-Gold}}]{Sayeed2023}
{Sayeed}, M., {Ness}, M.~K., {Montet}, B.~T., {et~al.} 2024, \apj, 964, 42

\bibitem[{{Singh} {et~al.}(2019){Singh}, {Reddy}, {Bharat Kumar}, \& {Antia}}]{Singh19}
{Singh}, R., {Reddy}, B.~E., {Bharat Kumar}, Y., \& {Antia}, H.~M. 2019, \apjl, 878, L21

\bibitem[{{Smiljanic, R.} {et~al.}(2009){Smiljanic, R.}, {Gauderon, R.}, {North, P.}, {Barbuy, B.}, {Charbonnel, C.}, \& {Mowlavi, N.}}]{Smiljanic2009}
{Smiljanic, R.}, {Gauderon, R.}, {North, P.}, {et~al.} 2009, A\&A, 502, 267

\bibitem[{{Sneden} {et~al.}(2022){Sneden}, {Af{\c{s}}ar}, {Bozkurt}, {Adam{\'o}w}, {Mallick}, {Reddy}, {Janowiecki}, {Mahadevan}, {Bowler}, {Hawkins}, {Lind}, {Dupree}, {Ninan}, {Nagarajan}, {Topcu}, {Froning}, {Bender}, {Terrien}, {Ramsey}, \& {Mace}}]{sneden22}
{Sneden}, C., {Af{\c{s}}ar}, M., {Bozkurt}, Z., {et~al.} 2022, \apj, 940, 12

\bibitem[{Sneden(1973)}]{sneden1973}
Sneden, C.~A. 1973, PhD thesis, The University of Texas at Austin, USA

\bibitem[{{Sun} {et~al.}(2022){Sun}, {Deliyannis}, {Twarog}, {Anthony-Twarog}, {Cummings}, \& {Steinhauer}}]{Sun2022}
{Sun}, Q., {Deliyannis}, C.~P., {Twarog}, B.~A., {et~al.} 2022, \mnras, 513, 5387

\bibitem[{{Tayar} {et~al.}(2015){Tayar}, {Ceillier}, {Garc{\'\i}a-Hern{\'a}ndez}, {Troup}, {Mathur}, {Garc{\'\i}a}, {Zamora}, {Johnson}, {Pinsonneault}, {M{\'e}sz{\'a}ros}, {Allende Prieto}, {Chaplin}, {Elsworth}, {Hekker}, {Nidever}, {Salabert}, {Schneider}, {Serenelli}, {Shetrone}, \& {Stello}}]{Tayar2015}
{Tayar}, J., {Ceillier}, T., {Garc{\'\i}a-Hern{\'a}ndez}, D.~A., {et~al.} 2015, \apj, 807, 82

\bibitem[{{Tsantaki} {et~al.}(2018){Tsantaki}, {Andreasen}, {Teixeira}, {Sousa}, {Santos}, {Delgado-Mena}, \& {Bruzual}}]{FASMA}
{Tsantaki}, M., {Andreasen}, D.~T., {Teixeira}, G.~D.~C., {et~al.} 2018, \mnras, 473, 5066

\bibitem[{{Tsantaki} {et~al.}(2023){Tsantaki}, {Delgado-Mena, E.}, {Bossini, D.}, {Sousa, S. G.}, {Pancino, E.}, \& {Martins, J. H. C.}}]{MariaLi}
{Tsantaki}, M., {Delgado-Mena, E.}, {Bossini, D.}, {et~al.} 2023, A\&A, 674, A157

\bibitem[{{Udry} {et~al.}(2000){Udry}, {Mayor}, {Naef}, {Pepe}, {Queloz}, {Santos}, {Burnet}, {Confino}, \& {Melo}}]{Udry2000}
{Udry}, S., {Mayor}, M., {Naef}, D., {et~al.} 2000, \aap, 356, 590

\bibitem[{{Xing} {et~al.}(2021){Xing}, {Li}, {Chang}, {Wang}, \& {Bai}}]{Xing2021}
{Xing}, L.-F., {Li}, Y.-C., {Chang}, L., {Wang}, C.-J., \& {Bai}, J.-M. 2021, \aap, 653, A28

\bibitem[{{Yana Galarza} {et~al.}(2019){Yana Galarza}, {Mel{\'e}ndez}, {Lorenzo-Oliveira}, {Valio}, {Reggiani}, {Carlos}, {Ponte}, {Spina}, {Haywood}, \& {Gandolfi}}]{yana-galarza19}
{Yana Galarza}, J., {Mel{\'e}ndez}, J., {Lorenzo-Oliveira}, D., {et~al.} 2019, \mnras, 490, L86

\bibitem[{{Zboril} {et~al.}(1997){Zboril}, {Byrne}, \& {Rolleston}}]{Zboril1997dwarfs}
{Zboril}, M., {Byrne}, P.~B., \& {Rolleston}, W.~R.~J.~R. 1997, \mnras, 284, 685

\bibitem[{{Zechmeister} \& {K{\"u}rster}(2009)}]{Zechmeister09}
{Zechmeister}, M. \& {K{\"u}rster}, M. 2009, \aap, 496, 577

\end{thebibliography}

\onecolumn
\begin{appendix}

\section{Additional tables}
\label{addtabs}
\begin{table*}[h!]
\centering
\caption{Summary of strong correlations found for the stars analysed in clusters IC4651, IC4756, NGC2345, NGC2423, NGC3114, NGC3532, NGC3680, NGC4349 and NGC5822.}
\resizebox{\textwidth}{!}{%
\begin{tabular}{@{}cccccccccc@{}}
Star          & $\rho_w$(RV,BIS) & $\rho_w$(RV,FWHM) &  $\rho_w$(RV,\teff)  & $\rho_w$(RV,\ha 1.6\AA) & $\rho_w$(RV,\ha 0.6\AA) & A(Li) [dex] & Mass (M$_\odot$) & $v\sin(i)$ (\kms)& n \\ \midrule
IC4651No10393 &               &          &          &               &               & -0.500                        & 1.801 & 1.870                          & 23 \\
IC4651No11453 &               &          &          &               &               & -0.007                        & 2.699 & 1.750                          & 33 \\
IC4651No12935 &               &          &          & $\times$      &               & 0.315                         & 1.800 & 0.000                          & 5  \\
IC4651No14527 &               &          &          &               &               & -0.500                        & 1.867 & 1.810                          & 31 \\
IC4651No6333  & $\times$      &          &          & $\times$      &               & \cellcolor[HTML]{DAEEF3}1.209 & 1.779 & 2.130                          & 5  \\
IC4651No7646  &               &          &          &               &               & -0.500                        & 2.438 & 0.000                          & 31 \\
IC4651No8540  &               &          &          &               &               & 0.715                         & 1.797 & 0.000                          & 27 \\
IC4651No9025  &               &          &          &               &               & 0.621                         & 1.815 & 0.000                          & 30 \\
IC4651No9122  &               &          &          &               &               & -0.500                        & 1.795 & 0.680                          & 74 \\
IC4651No9791  &               &          &          &               &               & \cellcolor[HTML]{DAEEF3}1.277 & 1.750 & 1.850                          & 33 \\\midrule
IC4756No14    &               & $\times$ & $\times$ &               &               & -0.016                        & 1.973 & 2.030                          & 5  \\
IC4756No38    & $\times$      &          &          &               &               & 0.783                         & 2.005 & 0.000                          & 7  \\
IC4756No44    & $\times$      &          &          &               & $\times$      & 0.823                         & 2.016 & 2.830                          & 6  \\
IC4756No52    &               &          &          & $\times$      &               & \cellcolor[HTML]{DAEEF3}1.447 & 2.272 & 2.870                          & 16 \\\midrule
NGC2345No50   &               & $\times$ &          & $\times\star$ & $\times\star$ & \cellcolor[HTML]{DAEEF3}0.977 & 5.843 & \cellcolor[HTML]{B1A0C7}5.270  & 26 \\\midrule
NGC2423No3    & $\times\star$ & $\star$  &          &               &               & \cellcolor[HTML]{DAEEF3}1.377 & 2.035 & 2.190                          & 60 \\
NGC2423No56   &               &          &          & $\times$      &               & 1.035                         & 1.991 & 0.000                          & 7  \\\midrule
NGC3114No150  & $\times$      &          &          &               &               & 1.340                         & 4.490 & \cellcolor[HTML]{B1A0C7}16.190 & 11 \\
NGC3114No170  & $\times$      & $\times$ &          & $\times$      & $\times$      & \cellcolor[HTML]{DAEEF3}1.503 & 3.981 & \cellcolor[HTML]{B1A0C7}11.320 & 6  \\
NGC3114No181  &               & $\times$ & $\times$ &               &               & 0.309                         & 4.022 & 3.950                          & 11 \\
NGC3114No223  &               &          & $\times$ &               &               & -0.500                        & 2.508 & 0.330                          & 8  \\
NGC3114No238  &               &          &          &               &               & 0.533                         & 3.807 & 4.090                          & 10 \\
NGC3114No262  & $\times$      &          &          &               &               & 1.190                         & 3.884 & \cellcolor[HTML]{B1A0C7}9.220  & 31 \\
NGC3114No283  & $\times$      &          &          &               &               & 1.415                         & 4.663 & \cellcolor[HTML]{B1A0C7}12.470 & 30 \\
NGC3114No6    &               & $\times$ &          &               &               & 1.092                         & 4.612 & \cellcolor[HTML]{B1A0C7}7.480  & 44 \\\midrule
NGC3532No100  &               &          &          &               &               & 0.952                         & 3.379 & 4.870                          & 19 \\
NGC3532No122  & $\times$      &          &          &               &               & 1.344                         & 2.873 & \cellcolor[HTML]{B1A0C7}9.570  & 20 \\
NGC3532No160  & $\times$      & $\times$ & $\times$ & $\times$      &               & 1.125                         & 3.149 & 4.780                          & 8  \\
NGC3532No19   &               &          &          &               &               & \cellcolor[HTML]{DAEEF3}1.443 & 3.218 & \cellcolor[HTML]{B1A0C7}5.050  & 46 \\
NGC3532No221  & $\times$      &          & $\times$ & $\times$      &               & 0.186                         & 4.818 & \cellcolor[HTML]{B1A0C7}5.890  & 22 \\
NGC3532No522  & $\times$      & $\times$ & $\times$ &               &               & 1.202                         & 4.412 & \cellcolor[HTML]{B1A0C7}12.480 & 6  \\
NGC3532No596  &               &          &          & $\times$      &               & 1.153                         & 3.126 & \cellcolor[HTML]{B1A0C7}7.210  & 31 \\
NGC3532No670  &               & $\star$  &          & $\star$       & $\times$      & \cellcolor[HTML]{DAEEF3}1.475 & 3.047 & 4.640                          & 29 \\\midrule
NGC3680No13   & $\times$      &          & $\times$ & $\times$      &               & \cellcolor[HTML]{DAEEF3}1.252 & 1.659 & 0.280                          & 7  \\
NGC3680No26   &               &          &          &               &               & 1.192                         & 1.704 & 0.960                          & 18 \\
NGC3680No34   &               &          &          &               &               & 0.246                         & 1.752 & 3.010                          & 12 \\
NGC3680No41   &               &          &          &               &               & 0.633                         & 1.641 & 0.000                          & 18 \\
NGC3680No44   &               & $\times$ &          & $\times$      & $\times$      & -0.500                        & 1.690 & 2.550                          & 6  \\
NGC3680No53   &               &          &          &               &               & \cellcolor[HTML]{DAEEF3}1.211 & 1.652 & 0.120                          & 18 \\\midrule
NGC4349No127  &               & $\star$  &          &               &               & \cellcolor[HTML]{DAEEF3}1.375 & 3.007 & 4.810                          & 56 \\
NGC4349No168  &               &          &          &               &               & 1.000                         & 3.358 & \cellcolor[HTML]{B1A0C7}5.510  & 35 \\
NGC4349No174  &               & $\times$ & $\times$ &               &               & 0.843                         & 3.003 & 3.190                          & 22 \\
NGC4349No203  &               & $\times$ &          &               &               & 1.105                         & 3.296 & \cellcolor[HTML]{B1A0C7}5.310  & 10 \\
NGC4349No5    &               &          &          &               &               & -0.500                        & 3.134 & \cellcolor[HTML]{B1A0C7}7.270  & 35 \\
NGC4349No9    & $\times$      &          &          &               & $\times$      & \cellcolor[HTML]{DAEEF3}1.262 & 3.005 & \cellcolor[HTML]{B1A0C7}9.580  & 24 \\\midrule
NGC5822No102  & $\times$      & $\times$ &          & $\times$      & $\times$      & \cellcolor[HTML]{DAEEF3}1.348 & 2.020 & \cellcolor[HTML]{B1A0C7}5.830  & 5  \\
NGC5822No1    &               &          &          &               &               & 0.126                         & 2.274 & 3.120                          & 8  \\
NGC5822No201  &               &          &          &               &               & 1.009                         & 2.422 & 1.600                          & 23 \\
NGC5822No240  &               &          &          &               &               & \cellcolor[HTML]{DAEEF3}1.401 & 2.071 & 2.830                          & 7  \\
NGC5822No316  & $\times$      &          & $\times$ & $\times$      & $\times$      & 0.636                         & 2.205 & 0.000                          & 7  \\
NGC5822No375  &               &          &          &               &               & -0.076                        & 2.142 & 2.190                          & 19 \\
NGC5822No443  &               & $\times$ &          &               &               & -0.138                        & 2.111 & 1.830                          & 8  \\
NGC5822No8    &               &          &          &               &               & 0.431                         & 2.262 & 0.360                          & 36 \\ \bottomrule
\end{tabular}}
\tablefoot{Each cross represents a correlation with BIS, FWHM, \teff\ or \ha\ $>0.4$. Highlighted in blue are every star in this part of the sample that can be considered Li-rich, and highlighted in purple are all stars with considerable rotation, that is, $v\sin(i)>5$ \kms. $\star$ indicates that a significant peak was found for the GLSP of the respective quantity which also appeared in RV. Also in this table are the values of the mass for each star (M) as well as the number of points (n) used to calculate the correlations.}
\label{summarypart1}
\end{table*}

\begin{table*}[h!]
\centering
\caption{Summary of strong correlations found for the stars analysed in cluster NGC6705.}
\resizebox{\textwidth}{!}{%
\begin{tabular}{@{}cccccccccc@{}}
Star          & $\rho_w$(RV,BIS) & $\rho_w$(RV,FWHM) &  $\rho_w$(RV,\teff)  & $\rho_w$(RV,\ha 1.6\AA) & $\rho_w$(RV,\ha 0.6\AA) & A(Li) [dex] & Mass (M$_\odot$) & $v\sin(i)$ (\kms) & n\\ \midrule
NGC6705No1090 &               &          &          &               &               & 1.380                         & 3.800 & \cellcolor[HTML]{B1A0C7}6.050  & 9  \\
NGC6705No1101 & $\star$       &          &          &               &               & \cellcolor[HTML]{DAEEF3}1.527 & 3.674 & \cellcolor[HTML]{B1A0C7}11.240 & 25 \\
NGC6705No1111 &               & $\times$ &          & $\times$      &               & 1.425                         & 3.618 & \cellcolor[HTML]{B1A0C7}7.110  & 10 \\
NGC6705No1117 &               &          &          & $\times$      & $\times$      & \cellcolor[HTML]{DAEEF3}1.567 & 3.521 & \cellcolor[HTML]{B1A0C7}9.440  & 10 \\
NGC6705No1145 & $\times$      & $\times$ &          &               &               & 1.385                         & 3.355 & \cellcolor[HTML]{B1A0C7}7.300  & 11 \\
NGC6705No1184 &               &          & $\times$ & $\times$      &               & 0.215                         & 3.448 & 4.420                          & 11 \\
NGC6705No1248 &               & $\times$ &          & $\times$      &               & \cellcolor[HTML]{DAEEF3}1.525 & 3.455 & \cellcolor[HTML]{B1A0C7}8.430  & 9  \\
NGC6705No1256 & $\times$      &          &          &               &               & \cellcolor[HTML]{DAEEF3}1.591 & 3.146 & 4.470                          & 14 \\
NGC6705No1286 & $\times$      &          &          &               & $\times$      & 1.400                         & 3.709 & \cellcolor[HTML]{B1A0C7}9.940  & 21 \\
NGC6705No1364 &               &          &          &               &               & \cellcolor[HTML]{DAEEF3}1.530 & 3.670 & \cellcolor[HTML]{B1A0C7}12.240 & 13 \\
NGC6705No136  &               &          & $\times$ &               & $\times$      & 1.082                         & 3.792 & 3.730                          & 7  \\
NGC6705No1423 &               &          &          &               &               & 1.455                         & 3.852 & \cellcolor[HTML]{B1A0C7}5.480  & 13 \\
NGC6705No1446 &               &          & $\times$ & $\times$      & $\times$      & 1.194                         & 3.451 & \cellcolor[HTML]{B1A0C7}7.140  & 6  \\
NGC6705No160  &               &          &          &               &               & 1.477                         & 3.740 & \cellcolor[HTML]{B1A0C7}6.950  & 22 \\
NGC6705No1625 & $\times$      &          &          &               & $\times$      & 1.088                         & 3.087 & 4.100                          & 19 \\
NGC6705No1658 &               &          &          &       &       & \cellcolor[HTML]{DAEEF3}1.512 & 3.910 & \cellcolor[HTML]{B1A0C7}7.290  & 28 \\
NGC6705No1837 & $\times$      & $\times$ & $\times$ &               &               & 1.495                         & 3.493 & \cellcolor[HTML]{B1A0C7}10.520 & 9  \\
NGC6705No2000 &               &          &          &               & $\times$      & 1.443                         & 4.040 & 4.760                          & 12 \\
NGC6705No320  &               & $\times$ &          & $\times$      & $\times$      & 1.492                         & 3.876 & \cellcolor[HTML]{B1A0C7}5.500  & 8  \\
NGC6705No411  &               & $\times$ &          & $\times$      & $\times$      & \cellcolor[HTML]{DAEEF3}1.573 & 3.193 & \cellcolor[HTML]{B1A0C7}6.470  & 8  \\
NGC6705No660  &               & $\times$ & $\times$ &               & $\times$      & \cellcolor[HTML]{DAEEF3}1.533 & 3.577 & \cellcolor[HTML]{B1A0C7}5.740  & 7  \\
NGC6705No669  &               &          &          &               & $\times$      & \cellcolor[HTML]{DAEEF3}1.511 & 3.451 & \cellcolor[HTML]{B1A0C7}8.930  & 6  \\
NGC6705No686  &               &          &          &               &               & 1.467                         & 3.727 & \cellcolor[HTML]{B1A0C7}9.480  & 6  \\
NGC6705No779  &               &          &          &               &               & \cellcolor[HTML]{DAEEF3}1.569 & 3.397 & \cellcolor[HTML]{B1A0C7}5.570  & 22 \\
NGC6705No816  &               &          & $\times$ &               & $\times$      & \cellcolor[HTML]{DAEEF3}1.533 & 3.956 & \cellcolor[HTML]{B1A0C7}8.770  & 6  \\
NGC6705No827  &               & $\times$ &          &               &               & -0.087                        & 4.040 & 4.800                          & 8  \\
NGC6705No916  & $\times$      &          &          &               &               & \cellcolor[HTML]{DAEEF3}1.532 & 3.745 & \cellcolor[HTML]{B1A0C7}9.700  & 9  \\
NGC6705No963  &               &          &          &               &               & \cellcolor[HTML]{DAEEF3}1.542 & 3.757 & \cellcolor[HTML]{B1A0C7}6.710  & 9  \\ \bottomrule
\end{tabular}}
\tablefoot{Each cross represents a correlation with BIS, FWHM, \teff\  or \ha\ $>0.4$. Highlighted in blue are every star in this part of the sample that can be considered Li-rich, and highlighted in purple are all stars with considerable rotation, that is, $v\sin(i)>5$ \kms. $\star$ indicates that a significant peak was found for the GLSP of the respective quantity which also appeared in RV. Also in this table are the values of the mass for each star (M) as well as the number of points (n) used to calculate the correlations.}
\label{summarypart2}
\end{table*}
\clearpage
\section{Correlation tables}
\label{Corrs}
\begin{table}[h!]
\centering
\caption{Correlations found for all of the stars in our sample.}
\resizebox{\textwidth}{10cm}{%
\begin{tabular}{@{}ccccccccccccccc@{}}
\toprule
Star          & $\rho_w$(RV,BIS)                 & p(BIS)   & $\rho_w$(RV,FWHM)                & p(FWHM)  & $\rho_w$(RV,\teff)               & p(\teff) & $\rho_w$(RV,\ha 1.6\AA)          & p(\ha 1.6\AA) & $\rho_w$(RV,\ha 0.6\AA)          & p(\ha 0.6\AA) & A(Li) [dex]                  & Mass (M$_\odot$) & $v\sin(i)$ (\kms)& n             \\ \midrule
IC4651No10393 & -0.12                         & 0.58              & -0.03                         & 0.89              & -0.11                         & 0.63          & -0.25                         & 0.26              & -0.32                         & 0.14              & -0.500                        & 1.801 & 1.870                          & 23 \\
IC4651No11453 & -0.09                         & 0.62              & 0.06                          & 0.75              & 0.17                          & 0.33          & 0.01                          & 0.95              & -0.001                        & 0.996             & -0.007                        & 2.699 & 1.750                          & 33 \\
IC4651No12935 & -0.02                         & 0.98              & 0.25                          & 0.68              & 0.11                          & 0.86          & \cellcolor[HTML]{F7F4CB}-0.44 & 0.45              & 0.07                          & 0.91              & 0.315                         & 1.800 & 0.000                          & 5  \\
IC4651No14527 & -0.02                         & 0.90              & 0.03                          & 0.88              & 0.23                          & 0.22          & -0.29                         & 0.12              & -0.34                         & 0.06              & -0.500                        & 1.867 & 1.810                          & 31 \\
IC4651No6333  & \cellcolor[HTML]{D8E4BC}-0.81 & 0.10              & 0.28                          & 0.64              & -0.13                         & 0.84          & \cellcolor[HTML]{D8E4BC}-0.65 & 0.23              & -0.23                         & 0.71              & \cellcolor[HTML]{DAEEF3}1.209 & 1.779 & 2.130                          & 5  \\
IC4651No7646  & -0.01                         & 0.95              & 0.29                          & 0.13              & 0.02                          & 0.90          & -0.13                         & 0.49              & -0.18                         & 0.34              & -0.500                        & 2.438 & 0.000                          & 31 \\
IC4651No8540  & -0.28                         & 0.15              & 0.24                          & 0.23              & -0.01                         & 0.97          & -0.24                         & 0.24              & 0.23                          & 0.25              & 0.715                         & 1.797 & 0.000                          & 27 \\
IC4651No9025  & -0.23                         & 0.24              & 0.09                          & 0.66              & -0.12                         & 0.55          & 0.20                          & 0.33              & -0.22                         & 0.26              & 0.621                         & 1.815 & 0.000                          & 30 \\
IC4651No9122  & 0.18                          & 0.13              & -0.12                         & 0.33              & -0.05                         & 0.66          & 0.10                          & 0.41              & 0.19                          & 0.11              & -0.500                        & 1.795 & 0.680                          & 74 \\
IC4651No9791  & -0.32                         & 0.07              & 0.03                          & 0.85              & -0.07                         & 0.70          & 0.07                          & 0.69              & 0.07                          & 0.71              & \cellcolor[HTML]{DAEEF3}1.277 & 1.750 & 1.850                          & 33 \\\midrule
IC4756No14    & -0.14                         & 0.82              & \cellcolor[HTML]{D8E4BC}0.69  & 0.19              & \cellcolor[HTML]{D8E4BC}-0.65 & 0.24          & 0.06                          & 0.92              & 0.25                          & 0.68              & -0.016                        & 1.973 & 2.030                          & 5  \\
IC4756No38    & \cellcolor[HTML]{F7F4CB}-0.42 & 0.26              & -0.10                         & 0.79              & 0.01                          & 0.98          & -0.29                         & 0.45              & 0.31                          & 0.42              & 0.783                         & 2.005 & 0.000                          & 7  \\
IC4756No44    & \cellcolor[HTML]{F7F4CB}0.46  & 0.35              & 0.06                          & 0.90              & -0.16                         & 0.77          & -0.23                         & 0.67              & \cellcolor[HTML]{F7F4CB}0.51  & 0.30              & 0.823                         & 2.016 & 2.830                          & 6  \\
IC4756No52   & 0.04                                   & 0.90              & -0.08                                  & 0.78              & -0.26                                 & 0.35     & \cellcolor[HTML]{D8E4BC}{0.92}  & \textbf{1.54E{-06}} & -0.08                                  & 0.77               & \cellcolor[HTML]{DAEEF3}1.447 & 2.272 & 2.870                          & 16 \\\midrule
NGC2345No50   & 0.31                          & 0.12              & \cellcolor[HTML]{F7F4CB}-0.51 & \textbf{0.01}     & 0.02                          & 0.91          & \cellcolor[HTML]{D8E4BC}0.60  & \textbf{1.07E-03}     & \cellcolor[HTML]{F7F4CB}0.56  & \textbf{2.86E-03} & \cellcolor[HTML]{DAEEF3}0.977 & 5.843 & \cellcolor[HTML]{B1A0C7}5.270  & 26 \\\midrule
NGC2423No3    & \cellcolor[HTML]{F7F4CB}0.56  & \textbf{3.40E-06} & 0.28                          & \textbf{0.03}     & 0.11                          & 0.40          & 0.03                          & 0.79              & -0.07                         & 0.61              & \cellcolor[HTML]{DAEEF3}1.377 & 2.035 & 2.190                          & 60 \\
NGC2423No56   & -0.11                         & 0.82              & 0.12                          & 0.79              & -0.07                         & 0.88          & \cellcolor[HTML]{D8E4BC}-0.91 & \textbf{4.38E-03} & 0.04                          & 0.93              & 1.035                         & 1.991 & 0.000                          & 7  \\\midrule
NGC3114No150  & \cellcolor[HTML]{F7F4CB}-0.52 & 0.10              & -0.06                         & 0.86              & 0.19                          & 0.58          & -0.17                         & 0.61              & -0.24                         & 0.47              & 1.340                         & 4.490 & \cellcolor[HTML]{B1A0C7}16.190 & 11 \\
NGC3114No170  & \cellcolor[HTML]{F7F4CB}-0.48 & 0.34              & \cellcolor[HTML]{F7F4CB}-0.54 & 0.27              & 0.25                          & 0.64          & \cellcolor[HTML]{D8E4BC}-0.62 & 0.19              & \cellcolor[HTML]{D8E4BC}-0.62 & 0.19              & \cellcolor[HTML]{DAEEF3}1.503 & 3.981 & \cellcolor[HTML]{B1A0C7}11.320 & 6  \\
NGC3114No181  & -0.23                         & 0.47              & \cellcolor[HTML]{D8E4BC}0.62  & \textbf{0.03}     & \cellcolor[HTML]{F7F4CB}-0.59 & \textbf{0.04} & 0.40                          & 0.20              & 0.26                          & 0.42              & 0.309                         & 4.022 & 3.950                          & 11 \\
NGC3114No223  & -0.19                         & 0.64              & -0.23                         & 0.59              & \cellcolor[HTML]{F7F4CB}-0.49 & 0.22          & 0.17                          & 0.69              & 0.07                          & 0.87              & -0.500                        & 2.508 & 0.330                          & 8  \\
NGC3114No238  & 0.26                          & 0.44              & 0.26                          & 0.44              & -0.01                         & 0.97          & 0.29                          & 0.38              & 0.15                          & 0.65              & 0.533                         & 3.807 & 4.090                          & 10 \\
NGC3114No262  & \cellcolor[HTML]{F7F4CB}-0.48 & \textbf{0.01}     & -0.03                         & 0.89              & 0.29                          & 0.11          & -0.04                         & 0.82              & 0.16                          & 0.39              & 1.190                         & 3.884 & \cellcolor[HTML]{B1A0C7}9.220  & 31 \\
NGC3114No283  & \cellcolor[HTML]{F7F4CB}0.49  & \textbf{0.01}     & 0.20                          & 0.30              & 0.09                          & 0.64          & 0.06                          & 0.74              & 0.27                          & 0.15              & 1.415                         & 4.663 & \cellcolor[HTML]{B1A0C7}12.470 & 30 \\
NGC3114No6    & -0.28                         & 0.06              & \cellcolor[HTML]{F7F4CB}-0.42 & \textbf{4.81E-03} & -0.18                         & 0.23          & 0.35                          & \textbf{0.02}     & 0.32                          & \textbf{0.03}     & 1.092                         & 4.612 & \cellcolor[HTML]{B1A0C7}7.480  & 44 \\\midrule
NGC3532No100  & 0.05                          & 0.85              & 0.20                          & 0.41              & -0.14                         & 0.56          & 0.19                          & 0.43              & 0.09                          & 0.71              & 0.952                         & 3.379 & 4.870                          & 19 \\
NGC3532No122  & \cellcolor[HTML]{D8E4BC}-0.60 & \textbf{4.94E-03} & 0.14                          & 0.56              & 0.09                          & 0.70          & 0.08                          & 0.74              & 0.14                          & 0.54              & 1.344                         & 2.873 & \cellcolor[HTML]{B1A0C7}9.570  & 20 \\
NGC3532No160  & \cellcolor[HTML]{D8E4BC}-0.75 & \textbf{0.03}     & \cellcolor[HTML]{F7F4CB}0.53  & 0.18              & \cellcolor[HTML]{D8E4BC}-0.65 & 0.08          & \cellcolor[HTML]{F7F4CB}0.56  & 0.15              & 0.19                          & 0.65              & 1.125                         & 3.149 & 4.780                          & 8  \\
NGC3532No19   & -0.12                         & 0.44              & -0.07                         & 0.66              & 0.11                          & 0.47          & 0.20                          & 0.17              & -0.11                         & 0.45              & \cellcolor[HTML]{DAEEF3}1.443 & 3.218 & \cellcolor[HTML]{B1A0C7}5.050  & 46 \\
NGC3532No221  & \cellcolor[HTML]{F7F4CB}-0.41 & \textbf{0.05}     & -0.27                         & 0.20              & \cellcolor[HTML]{D8E4BC}0.62  & \textbf{1.58E-03} & \cellcolor[HTML]{F7F4CB}-0.41 & 0.05              & 0.10                          & 0.66              & 0.186                         & 4.818 & \cellcolor[HTML]{B1A0C7}5.890  & 22 \\
NGC3532No522  & \cellcolor[HTML]{D8E4BC}-0.61 & 0.20              & \cellcolor[HTML]{D8E4BC}0.79  & 0.06              & \cellcolor[HTML]{F7F4CB}0.43  & 0.39          & 0.02                          & 0.97              & 0.10                          & 0.85              & 1.202                         & 4.412 & \cellcolor[HTML]{B1A0C7}12.480 & 6  \\
NGC3532No596  & 0.08                          & 0.66              & 0.01                          & 0.95              & 0.13                          & 0.49          & \cellcolor[HTML]{D8E4BC}-0.60 & \textbf{3.15E-04} & 0.10                          & 0.61              & 1.153                         & 3.126 & \cellcolor[HTML]{B1A0C7}7.210  & 31 \\
NGC3532No649  & -0.26                         & 0.08              & 0.10                          & 0.49              & 0.06                          & 0.69          & 0.13                          & 0.36              & 0.16                          & 0.26              & \cellcolor[HTML]{DAEEF3}3.271 & 2.167 & 0.000                          & 49 \\
NGC3532No670  & 0.18                          & 0.34              & -0.13                         & 0.49              & 0.09                          & 0.63          & 0.35                          & 0.06              & \cellcolor[HTML]{F7F4CB}0.42  & \textbf{0.02}     & \cellcolor[HTML]{DAEEF3}1.475 & 3.047 & 4.640                          & 29 \\\midrule
NGC3680No13   & \cellcolor[HTML]{D8E4BC}0.83  & \textbf{0.02}     & 0.08                          & 0.86              & \cellcolor[HTML]{F7F4CB}0.58  & 0.18          & \cellcolor[HTML]{F7F4CB}0.45  & 0.31              & 0.39                          & 0.38              & \cellcolor[HTML]{DAEEF3}1.252 & 1.659 & 0.280                          & 7  \\
NGC3680No26   & 0.02                          & 0.94              & 0.06                          & 0.82              & 0.17                          & 0.50          & 0.28                          & 0.27              & 0.35                          & 0.15              & 1.192                         & 1.704 & 0.960                          & 18 \\
NGC3680No34   & -0.13                         & 0.68              & -0.38                         & 0.22              & 0.02                          & 0.95          & -0.12                         & 0.71              & -0.15                         & 0.64              & 0.246                         & 1.752 & 3.010                          & 12 \\
NGC3680No41   & -0.03                         & 0.91              & -0.15                         & 0.55              & 0.20                          & 0.42          & 0.16                          & 0.53              & 0.01                          & 0.97              & 0.633                         & 1.641 & 0.000                          & 18 \\
NGC3680No44   & -0.02                         & 0.98              & \cellcolor[HTML]{D8E4BC}0.79  & 0.06              & 0.00                          & 0.99          & \cellcolor[HTML]{D8E4BC}0.77  & 0.08              & \cellcolor[HTML]{F7F4CB}0.46  & 0.36              & -0.500                        & 1.690 & 2.550                          & 6  \\
NGC3680No53   & -0.06                         & 0.81              & -0.04                         & 0.88              & 0.31                          & 0.20          & -0.03                         & 0.91              & -0.10                         & 0.70              & \cellcolor[HTML]{DAEEF3}1.211 & 1.652 & 0.120                          & 18 \\\midrule
NGC4349No127  & 0.04                          & 0.76              & -0.03                         & 0.80              & -0.07                         & 0.59          & 0.26                          & \textbf{0.05}     & 0.11                          & 0.44              & \cellcolor[HTML]{DAEEF3}1.375 & 3.007 & 4.810                          & 56 \\
NGC4349No168  & -0.17                         & 0.34              & 0.13                          & 0.46              & -0.04                         & 0.81          & -0.07                         & 0.69              & 0.03                          & 0.85              & 1.000                         & 3.358 & \cellcolor[HTML]{B1A0C7}5.510  & 35 \\
NGC4349No174  & -0.25                         & 0.25              & \cellcolor[HTML]{F7F4CB}-0.51 & \textbf{0.02}     & \cellcolor[HTML]{F7F4CB}0.50  & \textbf{0.02} & 0.07                          & 0.76              & 0.08                          & 0.72              & 0.843                         & 3.003 & 3.190                          & 22 \\
NGC4349No203  & 0.24                          & 0.50              & \cellcolor[HTML]{F7F4CB}-0.47 & 0.17              & 0.15                          & 0.69          & 0.27                          & 0.45              & 0.21                          & 0.56              & 1.105                         & 3.296 & \cellcolor[HTML]{B1A0C7}5.310  & 10 \\
NGC4349No5    & -0.24                         & 0.16              & -0.37                         & \textbf{0.03}     & 0.37                          & \textbf{0.03} & -0.04                         & 0.84              & -0.05                         & 0.79              & -0.500                        & 3.134 & \cellcolor[HTML]{B1A0C7}7.270  & 35 \\
NGC4349No9    & \cellcolor[HTML]{F7F4CB}-0.48 & \textbf{0.02}     & 0.37                          & 0.07              & -0.04                         & 0.86          & 0.13                          & 0.53              & \cellcolor[HTML]{F7F4CB}0.57  & \textbf{3.24E-03} & \cellcolor[HTML]{DAEEF3}1.262 & 3.005 & \cellcolor[HTML]{B1A0C7}9.580  & 24 \\\midrule
NGC5822No102  & \cellcolor[HTML]{D8E4BC}-0.84 & 0.07              & \cellcolor[HTML]{D8E4BC}1.00  & \textbf{1.20E-09} & 0.33                          & 0.59          & \cellcolor[HTML]{D8E4BC}-0.68 & 0.20              & \cellcolor[HTML]{D8E4BC}-0.68 & 0.21              & \cellcolor[HTML]{DAEEF3}1.348 & 2.020 & \cellcolor[HTML]{B1A0C7}5.830  & 5  \\
NGC5822No1    & -0.20                         & 0.63              & 0.10                          & 0.81              & -0.05                         & 0.91          & -0.02                         & 0.96              & 0.03                          & 0.95              & 0.126                         & 2.274 & 3.120                          & 8  \\
NGC5822No201  & -0.13                         & 0.56              & -0.21                         & 0.34              & -0.38                         & 0.07          & -0.12                         & 0.58              & -0.09                         & 0.68              & 1.009                         & 2.422 & 1.600                          & 23 \\
NGC5822No240  & -0.11                         & 0.82              & -0.26                         & 0.58              & -0.16                         & 0.73          & -0.20                         & 0.67              & -0.20                         & 0.67              & \cellcolor[HTML]{DAEEF3}1.401 & 2.071 & 2.830                          & 7  \\
NGC5822No316  & \cellcolor[HTML]{F7F4CB}-0.41 & 0.36              & 0.12                          & 0.80              & \cellcolor[HTML]{F7F4CB}-0.53 & 0.22          & \cellcolor[HTML]{F7F4CB}0.59  & 0.16              & \cellcolor[HTML]{F7F4CB}0.40  & 0.37              & 0.636                         & 2.205 & 0.000                          & 7  \\
NGC5822No375  & -0.21                         & 0.37              & -0.07                         & 0.76              & -0.35                         & 0.13          & -0.39                         & 0.09              & -0.19                         & 0.41              & -0.076                        & 2.142 & 2.190                          & 19 \\
NGC5822No443  & -0.35                         & 0.40              & \cellcolor[HTML]{F7F4CB}0.43  & 0.29              & -0.31                         & 0.46          & 0.26                          & 0.54              & -0.20                         & 0.64              & -0.138                        & 2.111 & 1.830                          & 8  \\
NGC5822No8    & -0.061                         & 0.724            & 0.151                          & 0.379            & -0.100                         & 0.563            & 0.119                        & 0.490            & 0.042                         & 0.807          & 0.431                         & 2.262 & 0.360                          & 36 \\\midrule
NGC6705No1090 & 0.01                          & 0.98              & 0.23                          & 0.56              & -0.08                         & 0.84          & -0.08                         & 0.84              & 0.11                          & 0.78              & 1.380                         & 3.800 & \cellcolor[HTML]{B1A0C7}6.050  & 9  \\
NGC6705No1101 & -0.40                         & \textbf{0.05}     & 0.22                          & 0.29              & 0.12                          & 0.57          & -0.15                         & 0.46              & 0.13                          & 0.53              & \cellcolor[HTML]{DAEEF3}1.527 & 3.674 & \cellcolor[HTML]{B1A0C7}11.240 & 25 \\
NGC6705No1111 & -0.28                         & 0.43              & \cellcolor[HTML]{F7F4CB}-0.48 & 0.16              & 0.27                          & 0.45          & \cellcolor[HTML]{F7F4CB}0.55  & 0.10              & 0.02                          & 0.95              & 1.425                         & 3.618 & \cellcolor[HTML]{B1A0C7}7.110  & 10 \\
NGC6705No1117 & -0.25                         & 0.49              & -0.13                         & 0.72              & 0.22                          & 0.54          & \cellcolor[HTML]{F7F4CB}-0.57 & 0.09              & \cellcolor[HTML]{F7F4CB}-0.47 & 0.17              & \cellcolor[HTML]{DAEEF3}1.567 & 3.521 & \cellcolor[HTML]{B1A0C7}9.440  & 10 \\
NGC6705No1145 & \cellcolor[HTML]{F7F4CB}-0.56 & 0.07              & \cellcolor[HTML]{D8E4BC}0.68  & \textbf{0.02}     & 0.39                          & 0.23          & -0.02                         & 0.96              & 0.28                          & 0.40              & 1.385                         & 3.355 & \cellcolor[HTML]{B1A0C7}7.300  & 11 \\
NGC6705No1184 & -0.07                         & 0.84              & 0.14                          & 0.69              & \cellcolor[HTML]{F7F4CB}0.43  & 0.19          & \cellcolor[HTML]{F7F4CB}-0.50 & 0.11              & -0.25                         & 0.45              & 0.215                         & 3.448 & 4.420                          & 11 \\
NGC6705No1248 & 0.09                          & 0.82              & \cellcolor[HTML]{F7F4CB}0.44  & 0.24              & 0.27                          & 0.48          & \cellcolor[HTML]{F7F4CB}-0.50 & 0.17              & -0.29                         & 0.45              & \cellcolor[HTML]{DAEEF3}1.525 & 3.455 & \cellcolor[HTML]{B1A0C7}8.430  & 9  \\
NGC6705No1256 & \cellcolor[HTML]{F7F4CB}0.43  & 0.12              & -0.15                         & 0.61              & 0.28                          & 0.34          & 0.11                          & 0.71              & -0.32                         & 0.27              & \cellcolor[HTML]{DAEEF3}1.591 & 3.146 & 4.470                          & 14 \\
NGC6705No1286 & \cellcolor[HTML]{D8E4BC}-0.71 & \textbf{3.99E-04} & -0.35                         & 0.13              & 0.11                          & 0.63          & 0.20                          & 0.39              & \cellcolor[HTML]{D8E4BC}0.67  & \textbf{1.32E-03} & 1.400                         & 3.709 & \cellcolor[HTML]{B1A0C7}9.940  & 21 \\
NGC6705No1364 & -0.11                         & 0.73              & 0.31                          & 0.30              & 0.05                          & 0.87          & 0.26                          & 0.38              & 0.32                          & 0.28              & \cellcolor[HTML]{DAEEF3}1.530 & 3.670 & \cellcolor[HTML]{B1A0C7}12.240 & 13 \\
NGC6705No136  & 0.05                          & 0.91              & 0.10                          & 0.83              & \cellcolor[HTML]{F7F4CB}0.47  & 0.29          & -0.17                         & 0.71              & \cellcolor[HTML]{F7F4CB}-0.44 & 0.32              & 1.082                         & 3.792 & 3.730                          & 7  \\
NGC6705No1423 & -0.31                         & 0.31              & 0.08                          & 0.79              & 0.15                          & 0.63          & 0.25                          & 0.41              & -0.15                         & 0.63              & 1.455                         & 3.852 & \cellcolor[HTML]{B1A0C7}5.480  & 13 \\
NGC6705No1446 & -0.02                         & 0.97              & 0.34                          & 0.51              & \cellcolor[HTML]{D8E4BC}0.67  & 0.14          & \cellcolor[HTML]{F7F4CB}-0.58 & 0.23              & \cellcolor[HTML]{D8E4BC}0.83  & \textbf{0.04}     & 1.194                         & 3.451 & \cellcolor[HTML]{B1A0C7}7.140  & 6  \\
NGC6705No160  & -0.29                         & 0.19              & 0.29                          & 0.19              & 0.22                          & 0.33          & -0.11                         & 0.62              & 0.11                          & 0.63              & 1.477                         & 3.740 & \cellcolor[HTML]{B1A0C7}6.950  & 22 \\
NGC6705No1625 & \cellcolor[HTML]{F7F4CB}-0.44 & 0.06              & -0.11                         & 0.66              & 0.25                          & 0.31          & -0.20                         & 0.40              & \cellcolor[HTML]{F7F4CB}0.41  & 0.08              & 1.088                         & 3.087 & 4.100                          & 19 \\
NGC6705No1658 & -0.26                         & 0.19              & -0.06                         & 0.76              & 0.07                          & 0.72          & -0.10                         & 0.63              & 0.13                          & 0.50              & \cellcolor[HTML]{DAEEF3}1.512 & 3.910 & \cellcolor[HTML]{B1A0C7}7.290  & 28 \\
NGC6705No1837 & \cellcolor[HTML]{D8E4BC}-0.77 & \textbf{0.01}     & \cellcolor[HTML]{F7F4CB}0.52  & 0.15              & \cellcolor[HTML]{D8E4BC}0.61  & 0.08          & -0.03                         & 0.94              & -0.06                         & 0.88              & 1.495                         & 3.493 & \cellcolor[HTML]{B1A0C7}10.520 & 9  \\
NGC6705No2000 & -0.13                         & 0.70              & -0.05                         & 0.88              & 0.30                          & 0.35          & 0.29                          & 0.36              & \cellcolor[HTML]{F7F4CB}-0.44 & 0.16              & 1.443                         & 4.040 & 4.760                          & 12 \\
NGC6705No320  & 0.01                          & 0.98              & \cellcolor[HTML]{F7F4CB}-0.44 & 0.27              & -0.21                         & 0.62          & \cellcolor[HTML]{F7F4CB}0.51  & 0.20              & \cellcolor[HTML]{F7F4CB}0.60  & 0.12              & 1.492                         & 3.876 & \cellcolor[HTML]{B1A0C7}5.500  & 8  \\
NGC6705No411  & 0.07                          & 0.86              & \cellcolor[HTML]{D8E4BC}0.61  & 0.11              & -0.10                         & 0.82          & \cellcolor[HTML]{F7F4CB}-0.47 & 0.24              & \cellcolor[HTML]{F7F4CB}0.52  & 0.18              & \cellcolor[HTML]{DAEEF3}1.573 & 3.193 & \cellcolor[HTML]{B1A0C7}6.470  & 8  \\
NGC6705No660  & -0.21                         & 0.65              & \cellcolor[HTML]{D8E4BC}-0.62 & 0.14              & \cellcolor[HTML]{F7F4CB}0.47  & 0.29          & -0.11                         & 0.82              & \cellcolor[HTML]{F7F4CB}-0.43 & 0.34              & \cellcolor[HTML]{DAEEF3}1.533 & 3.577 & \cellcolor[HTML]{B1A0C7}5.740  & 7  \\
NGC6705No669  & 0.38                          & 0.46              & -0.25                         & 0.64              & 0.13                          & 0.80          & 0.14                          & 0.79              & \cellcolor[HTML]{F7F4CB}0.48  & 0.34              & \cellcolor[HTML]{DAEEF3}1.511 & 3.451 & \cellcolor[HTML]{B1A0C7}8.930  & 6  \\
NGC6705No686  & -0.05                         & 0.93              & 0.35                          & 0.50              & 0.15                          & 0.78          & -0.07                         & 0.89              & -0.36                         & 0.48              & 1.467                         & 3.727 & \cellcolor[HTML]{B1A0C7}9.480  & 6  \\
NGC6705No779  & 0.34                          & 0.13              & -0.32                         & 0.15              & -0.18                         & 0.43          & 0.18                          & 0.43              & -0.05                         & 0.83              & \cellcolor[HTML]{DAEEF3}1.569 & 3.397 & \cellcolor[HTML]{B1A0C7}5.570  & 22 \\
NGC6705No816  & -0.19                         & 0.72              & 0.10                          & 0.85              & \cellcolor[HTML]{F7F4CB}0.41  & 0.41          & -0.05                         & 0.93              & \cellcolor[HTML]{F7F4CB}-0.58 & 0.23              & \cellcolor[HTML]{DAEEF3}1.533 & 3.956 & \cellcolor[HTML]{B1A0C7}8.770  & 6  \\
NGC6705No827  & 0.06                          & 0.89              & \cellcolor[HTML]{F7F4CB}-0.54 & 0.22              & 0.23                          & 0.62          & 0.24                          & 0.61              & -0.03                         & 0.95              & -0.087                        & 4.040 & 4.800                          & 8  \\
NGC6705No916  & \cellcolor[HTML]{F7F4CB}-0.58 & 0.13              & -0.09                         & 0.83              & 0.12                          & 0.77          & -0.07                         & 0.86              & 0.21                          & 0.62              & \cellcolor[HTML]{DAEEF3}1.532 & 3.745 & \cellcolor[HTML]{B1A0C7}9.700  & 9  \\
NGC6705No963  & -0.28                         & 0.47              & 0.27                          & 0.48              & 0.03                          & 0.93          & -0.21                         & 0.60              & 0.14                          & 0.71              & \cellcolor[HTML]{DAEEF3}1.542 & 3.757 & \cellcolor[HTML]{B1A0C7}6.710  & 9 

\\ \bottomrule
\end{tabular}%
}
\tablefoot{Correlations determined for all of the stars in our sample. Highlighted in yellow are all strong correlations ($>$0.4), in green are all the very strong correlations ($>$0.6). For every correlation that is calculated, a p-value comes associated to it. Correlations are deemed significant with a 95\% confidence level if their respective p-value is bellow 0.05. Highlighted in blue are all Li-rich stars, and highlighted in purple are all stars with a considerable rotation ($v\sin(i)>5$ \kms). The number of points used to calculate the correlations (n) and stellar mass in units of solar mass (M$_\odot$) are also provided.}
\end{table}
\clearpage
\section{Methods for data analysis}
\label{data_collect}

\begin{figure}[h!]
    \centering
    \includegraphics[width=\textwidth]{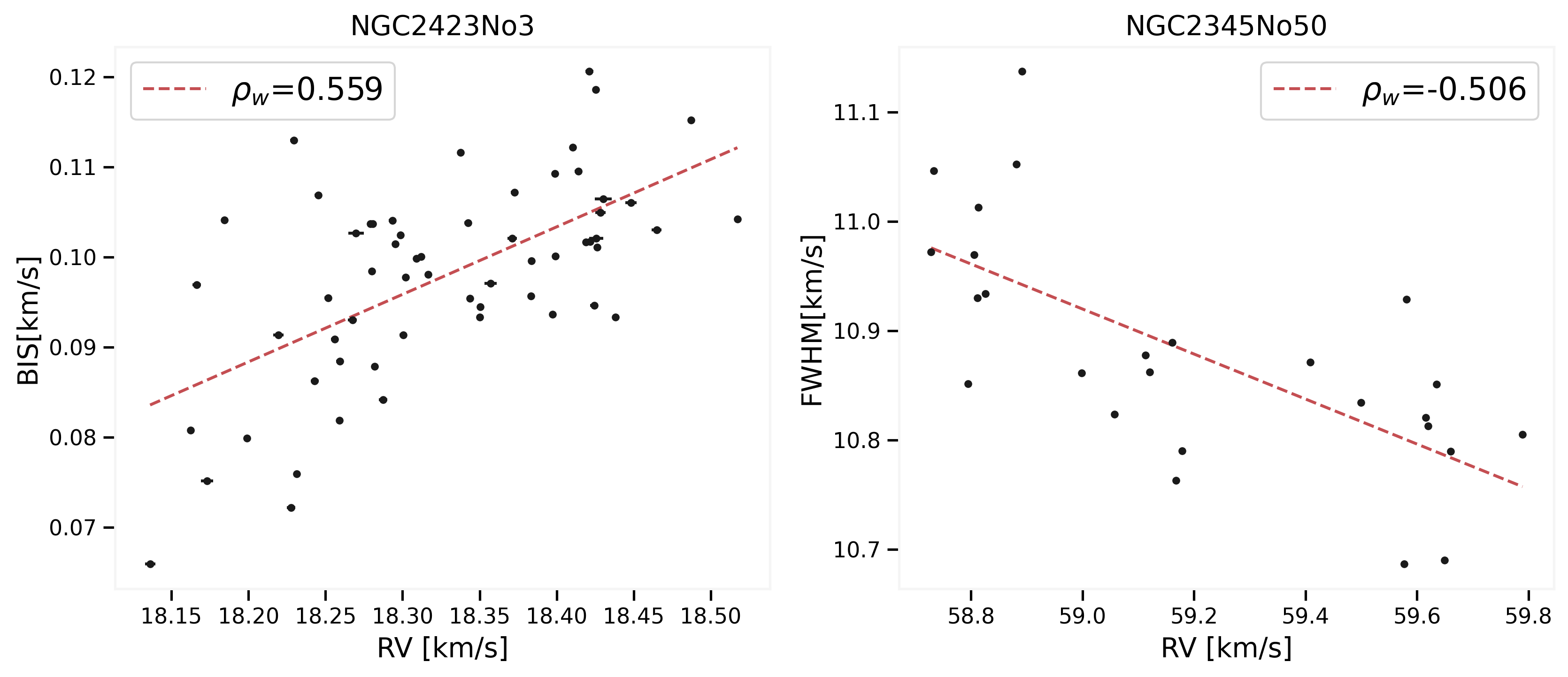}
    \caption{Example of strong correlations found for stars NGC2423No3 (left) and NGC2345No50 (right). NGC2423No3 shows a strong positive correlation between RV and BIS, while NGC23545No50 shows a strong negative correlation between RV and FWHM.}
    \label{fig:exmpl_corrs}
\end{figure}

\begin{figure}[h!]
    \centering
    \includegraphics[width=\textwidth]{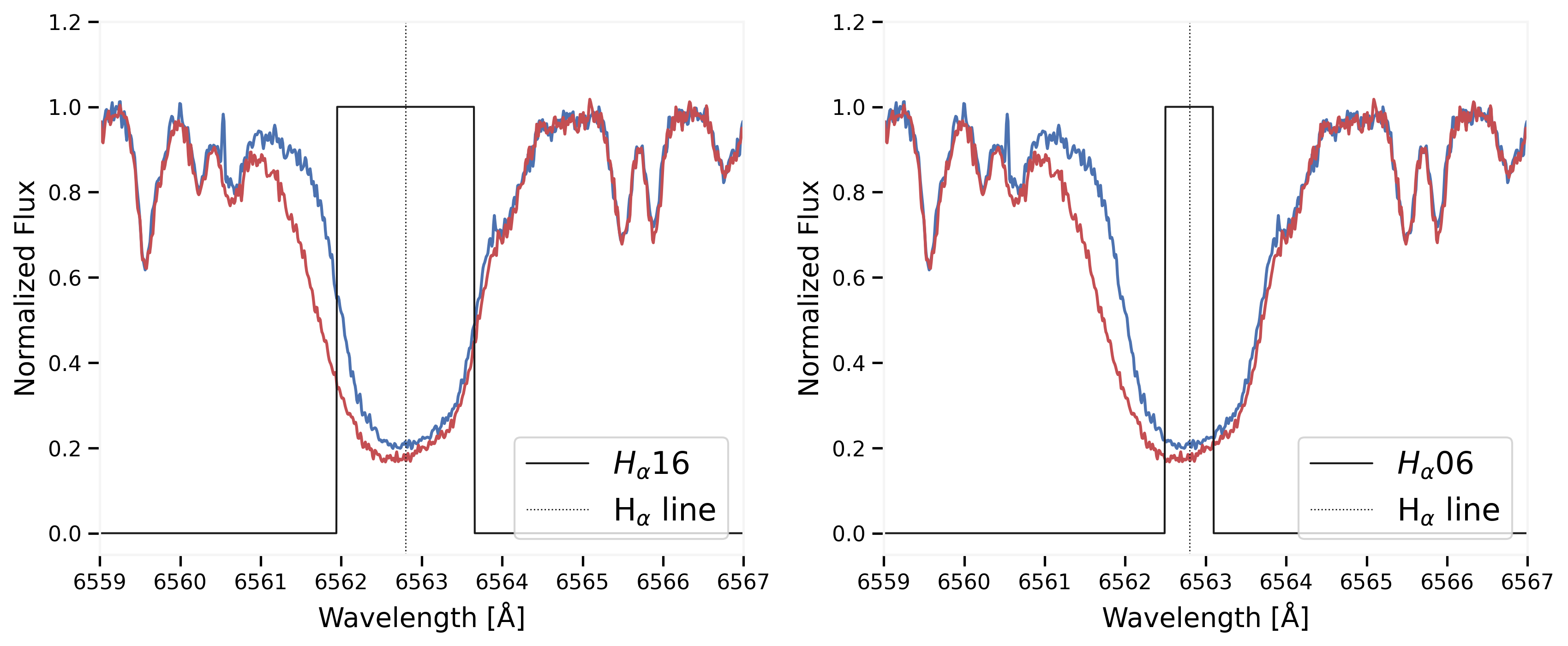}
    \caption{{Example of the \ha\ line profile difference of two spectra obtained at different times for star NGC2345No50 (depicted by the blue and red lines). In black are represented the two different bandpasses (\hal\ on the left and \has\ on the right), around the \ha\ line centre, used to compute the \ha\ index for the individual spectra of the stars with ACTIN 2.}}
    \label{fig:haband}
\end{figure}

\clearpage
\section{Additional relations}
\label{SSR}

\begin{figure}[h!]
\centering
    \includegraphics[scale=0.535]{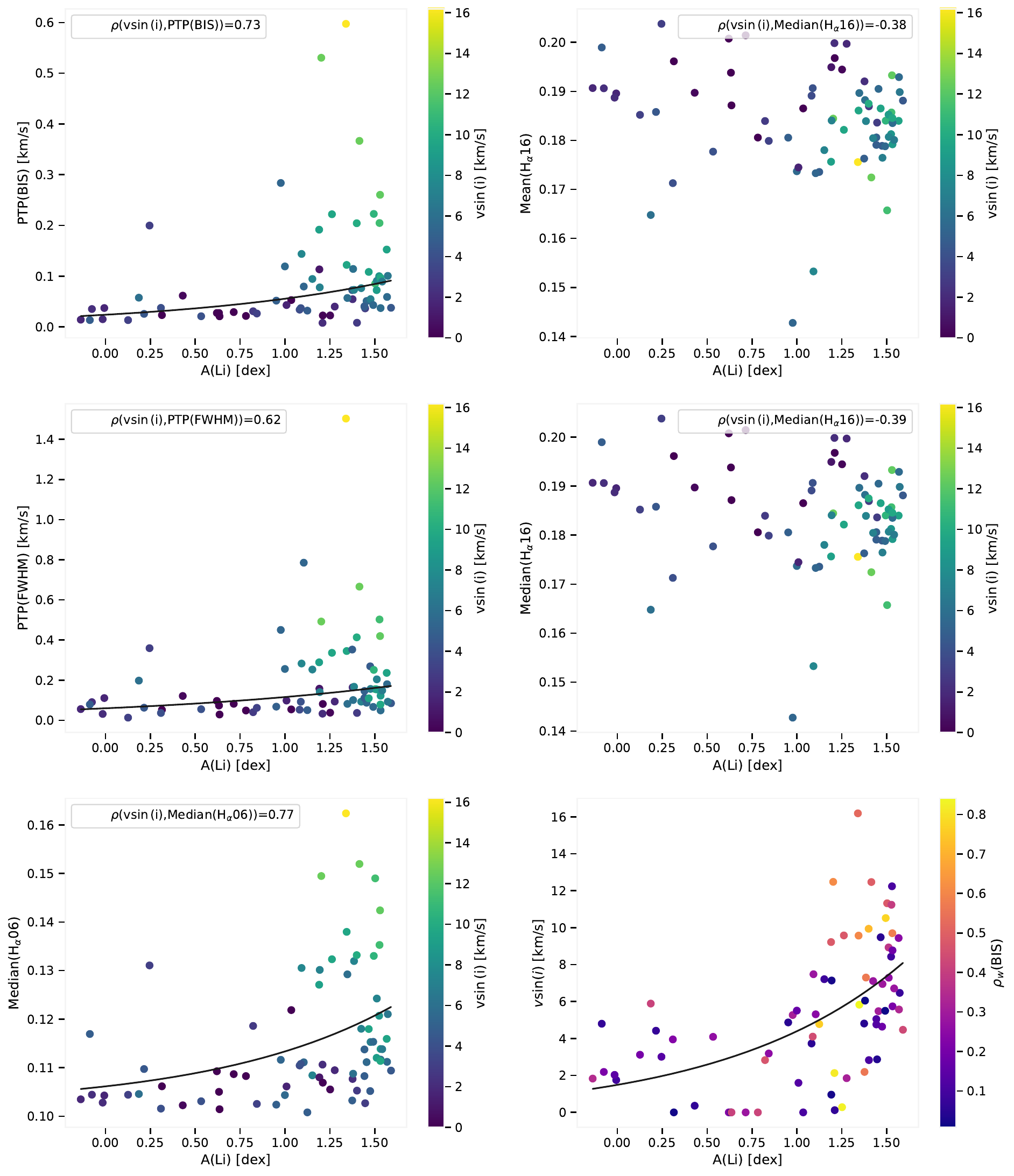}
    \caption{\textbf{Left:} Exponential relation found between the PTP of BIS data for each star and A(Li) [top panel]; the PTP of FWHM data for each star and A(Li) [middle panel]; the median of \has\ data for each star and A(Li) [bottom panel]. The data points are coloured according to the $v\sin(i)$ of each star. The black line is the line of best fit. In these plots we also show the Pearson Correlation Coefficients ($\rho$) between the activity proxies used and $v\sin(i)$, all of them being very strong ($>0.6$). \textbf{Right:} No relation found between the mean and median of \hal\ data for each star and A(Li) [top and middle panels, respectively]. Here again the data points are coloured according to the $v\sin(i)$ of each star and $\rho$ between the activity proxies and $v\sin(i)$ is also shown, being weak in both cases. In the bottom panel we show the exponential relation found between $v\sin(i)$ and A(Li). The data points are coloured according to the strength of the correlations of $\rho_w$(BIS) to demonstrate that the majority of correlations found for this indicator appear for moderate and fast rotators ($v\sin(i)>$5 km/s). The black line is the line of best fit.}
    \label{LiVsPTP}
\end{figure}

\clearpage

\section{Generalized Lomb-Scargle periodograms}
\label{LS}
The remainder of the GLPs mentioned in the text can be found in \cite{Elisa2023}.

\begin{figure}[h!]
    \begin{subfigure}{0.48\textwidth}
    \centering
    \includegraphics[width=\textwidth]{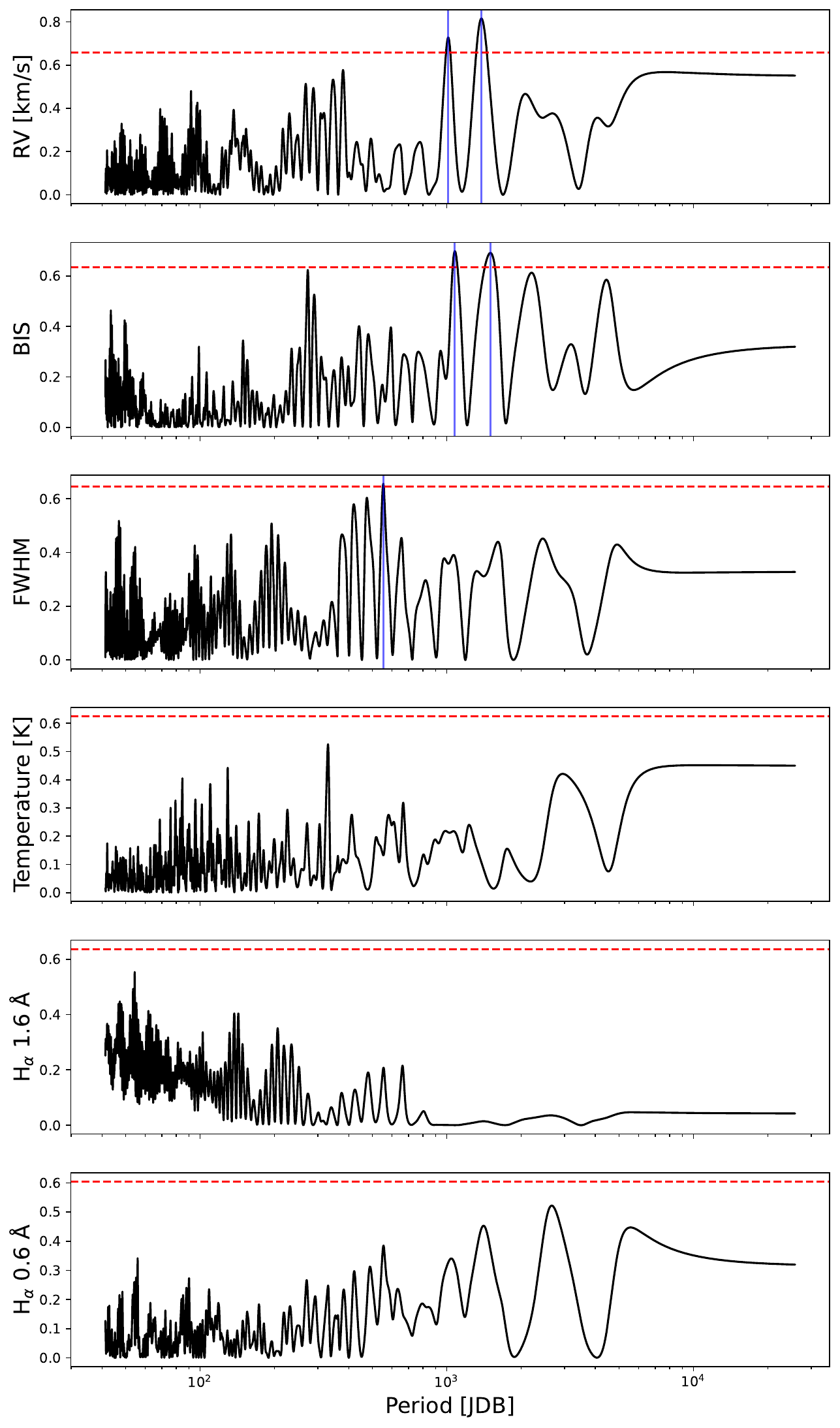}
    \caption{Generalised Lomb-Scargle periodograms pertaining to star \textit{NGC6705No1101} for BIS, FWHM, \teff, and \ha. The red line indicates a false alarm probability level of 1\%. Significant peaks are observed in the RV and BIS GLSPs at approximately the same period. There is also a significant peak observed in FWHM not related to those observed in RV.}
  \end{subfigure}
\end{figure}

\end{appendix}

\end{document}